\documentclass[showpacs,showkeys,preprintnumbers,amsmath,amssymb,floatfix,superscriptaddress]{revtex4}

\usepackage{graphicx}
\usepackage{dcolumn}
\usepackage{bm}

\begin{document}

\preprint{\em{Revised version published in Physica A {\bf 352}, 131 (2005)}}

\title{Electrostatic Interactions in Strongly-Coupled Soft Matter}

\author{Ali Naji}
    \email{naji@ph.tum.de}
 \affiliation{Physics Department, Technical University of Munich, James Franck St., \\D-85748 Garching, Germany.}       
 \affiliation{Physics Department, Ludwig Maximilian University, Theresienstr. 37, \\D-80333 Munich, Germany.}
\author{Swetlana Jungblut}
\affiliation{Physics Department, Ludwig Maximilian University, Theresienstr.  37, \\D-80333 Munich, Germany.}
\author{Andr\'e G. Moreira}
 \affiliation{Polymer Physics, BASF Aktiengesellschaft, 67056 Ludwigshafen, Germany.}
\author{Roland R. Netz}
  \affiliation{Physics Department, Technical University Munich, James Franck Str., \\D-85748 Garching, Germany.}       
 \affiliation{Physics Department, Ludwig Maximilian University, Theresienstr. 37, \\D-80333 Munich, Germany.}


\begin{abstract}
     Charged soft-matter systems--such as colloidal dispersions and charged polymers--are 
     dominated by attractive forces between 
    constituent  like-charged particles when neutralizing
     counterions of high charge valency are introduced. 
     Such counter-intuitive effects indicate strong electrostatic coupling 
     between like-charged particles, which essentially results from electrostatic correlations
     among counterions residing near particle surfaces.
     In this paper, the attraction mechanism and the structure of counterionic correlations are discussed 
     in the limit of strong coupling based on recent numerical and analytical investigations
     and for various geometries (planar, spherical and cylindrical) of charged objects.
\\
\end{abstract}

\maketitle


\section{Introduction}
\label{sec:intro}

Electrostatic interactions and processes involving electric charges appear ubiquitously 
in biological and soft-matter systems. Electric charges make materials water-soluble and lead to
many important technological and biological 
applications. In industry, for example, the stability of colloidal dispersions
is often a desirable property as in the case of paints and food emulsions such as milk. One way 
to stabilize colloidal suspensions against coagulation or flocculation
(occurring due to attractive van-der-Waals forces) is to generate long-range repulsive interactions
between constituent colloidal particles by imparting permanent like-charges 
to these particles \cite{VO,Israelachvili,Hunter}, or by grafting charged polymeric chains to their surfaces  
(forming hairy particles or polymeric brushes) \cite{Napper}. Charged polymers, or the so-called
{\em polyelectrolytes}, and their synthesis have also attracted a lot of attention \cite{Polyelec}
since, due to their water-solubility, they offer a useful option in design and processing 
of non-toxic environmentally-friendly materials.  
In biology, on the other hand, electrostatic effects emerge in many striking examples
such as the DNA-packaging process in the cell nucleus. 
In each human cell, a total length of about
$2$m of DNA--which bears a total negative charge of about $10^{10}e$ ({\em i.e.} one elementary charge, 
$e$, per $1.7$\AA)--is stored inside the cell nucleus with a diameter of less than $10\mu$m. 
This storage process involves a hierarchical structure on the lowest level of which, 
short segments of DNA are tightly wrapped around positively charged proteins of 
few-nanometer size (the so-called Histones). It is shown experimentally \cite{Yager} that such a tightly 
wrapped state is only stable for intermediate, physiological salt concentrations, at which an
optimal balance between self-repulsion of DNA segments and the DNA-Histone attraction is achieved.

Charged macromolecules (macroions) in solution are always surrounded by neutralizing counterions, 
and also in general by coions. Counterions form electrostatically-bound 
clouds in the proximity of macroions
and in many cases, predominantly determine the electric properties of charged solutions
\cite{Hunter}. In particular, counterions can alter the effective interaction between like-charged macroions, 
and may generate a dominant electrostatic attraction between them in certain physical conditions \cite{Khan,Wennerstroem91,Quirk,Bloom,Kekicheff93,Delsanti,Della,Dubois,Strey98,Tang,Tang03,Angelini03,Guld84,Svensson,Bratko86,Guld86,Wood88,Valleau,Kjellander92,Lyub95,Gron97,Gron98,Wu,AllahyarovPRL,Stevens99,LinsePRL,Linse00,Hribar,Messina00,Allahyarov00,AndreEPL,AndrePRL,AndreEPJE,Deserno03,Naji_epl04,Lee04,Kjellander84,Oosawa,Oosawa_book,Attard-etal,Attard-etal2,Podgornik89,Barrat,Pincus98,Podgornik98,Ha, Safran99, Kardar99,Netz-orland,Ha01,Lau02,Gonz01,Stevens90,Diehl99,Rouzina96,Korny,Arenzon99,Shklovs99a,Shklovs99,Shklovs02,Diehl01,Lau01,Netz01,Levin02,Naji_epje04,PhysToday,Rouzina98,Golestan99,Levin99,Shklo00,Belloni95,Manning97}.
 Like-charge attraction 
manifests itself in a number of famous examples, namely, the
condensation of DNA molecules \cite{Bloom}, bundle formation of  stiff polymers \cite{Tang}
and aggregation of  colloidal particles \cite{LinsePRL,Linse00,Hribar}.
Interestingly, such an attraction emerges only in {\em strongly-coupled} 
systems, {\em e.g.} when macroions are highly charged 
(with surface charge densities up to $1e/$nm$^2$ as in the DNA system), 
neutralizing counterions are multivalent, or when the temperature or the dielectric constant of medium
is low. For instance, the DNA condensation process,  in which long DNA molecules condense into a
tightly packed, circumferentially wound torus, is
observed in experiments where multivalent counterions (such as trivalent spermidine ions)
are introduced  \cite{Bloom}. 
A similar trend has also been found in numerous numerical simulations 
of like-charged membranes, colloids and polymers \cite{Guld84,Svensson,Bratko86,Guld86,Wood88,Valleau,Kjellander92,Lyub95,Gron97,Gron98,Wu,AllahyarovPRL,Stevens99,LinsePRL,Linse00,Hribar,Messina00,Allahyarov00,AndreEPL,AndrePRL,AndreEPJE,Deserno03,Naji_epl04,Lee04}, where highly charged macroions are found to form closely-packed bound states due to attractive forces of
electrostatic origin. These attractive forces are of typically large strength compared to the 
usual van-der-Waals attraction and may have significant practical implications where, for instance, 
multivalent counterions are present.  

In the weak-coupling condition, {\em e.g.} when surface charge densities and charge valency of counterions
are small, macroions are found to repel each other. In fact, from a theoretical point of view, 
one can argue that weakly-coupled systems should be described well by 
the mean-field approximation. Therefore, 
the effective interaction between macroions may be calculated using the mean-field
solution for the electrostatic potential field in space, which follows from
the so-called Poisson-Boltzmann (PB) equation that has been widely studied in the past 
\cite{VO,Israelachvili,Hunter}. It is known that the Poisson-Boltzmann theory--despite
its success in many applications--can only yield repulsion between like-charged 
macroions \cite{Neu,Sader,Trizac}.  

The main scenarios which are put forward to explain the phenomenon of like-charge attraction 
have gone beyond the mean-field level by demonstrating that this phenomenon 
can be reproduced quantitatively by inclusion of electrostatic correlations
\cite{Guld84,Svensson,Bratko86,Guld86,Wood88,Valleau,Kjellander92,Lyub95,Gron97,Gron98,Wu,AllahyarovPRL,Stevens99,LinsePRL,Linse00,Hribar,Messina00,Allahyarov00,AndreEPL,AndrePRL,AndreEPJE,Deserno03,Naji_epl04,Lee04,Kjellander84,Oosawa,Oosawa_book,Attard-etal,Attard-etal2,Podgornik89,Barrat,Pincus98,Podgornik98,Ha, Safran99, Kardar99,Netz-orland,Ha01,Lau02,Gonz01,Stevens90,Diehl99,Rouzina96,Korny,Arenzon99,Shklovs99a,Shklovs99,Shklovs02,Diehl01,Lau01,Netz01,Levin02,Naji_epje04,PhysToday,Rouzina98,Golestan99,Levin99,Shklo00}. 
These correlations are systematically
neglected on the mean-field level but become enhanced 
in strong-coupling conditions mentioned above. Recent theoretical attempts to incorporate ionic 
correlations include integral-equation methods \cite{Kjellander92,Kjellander84}, perturbative improvement 
of the mean-field theory including Gaussian-fluctuations theories \cite{Oosawa,Oosawa_book,Attard-etal,Attard-etal2,Podgornik89,Barrat,Pincus98,Podgornik98,Ha, Safran99, Kardar99,Netz-orland,Ha01,Lau02}, 
and local density functional theory \cite{Stevens90,Diehl99}, which 
compare well with numerical simulations and all exhibit attraction. 
These methods are mostly applicable 
for large separations between macroions or in the regime of low coupling
strengths (the so-called {\em high-temperature} regime), and can not characterize
the closely-packed bound state between like-charged macroions. 
An alternative approach was triggered by 
Rouzina and Bloomfield \cite{Rouzina96} with the insight that counterions form 
two-dimensional highly-correlated layers 
at macroionic surfaces for high coupling strengths (the so-called {\em low-temperature}
regime). Such structural correlations give rise to attractive forces  \cite{Korny,Arenzon99,Shklovs99a,Shklovs99,Shklovs02,Diehl01,Lau01,Netz01,Levin02,Naji_epje04}, which mainly result from 
energetic origins and can account for the closely-packed bound state of like-charged macroions
\cite{AndrePRL,AndreEPJE,Naji_epl04,Korny,Arenzon99,Shklovs99a,Shklovs99,Shklovs02,Diehl01,Lau01,Netz01,Levin02,Naji_epje04}.

In this paper, we shall chiefly consider this latter regime of strong coupling. 
First we shall briefly discuss the regime of parameters where 
electrostatic correlations are expected to be small or large in terms of the 
main length scales in a charged system (Section \ref{sec:scales}). 
In Sections \ref{sec:scales}-\ref{sec:twowalls}, we shall focus 
on the system of counterions at one or two planar charged walls to 
demonstrate the gross physical picture in both weak-coupling (or mean-field)
and strong-coupling regimes. The role of curvature and its main 
physical consequences are considered in Sections \ref{sec:curve}-\ref{sec:rods}. As we shall see,
the so-called counterion-condensation process occurring at curved surfaces
plays an important role in determining the attraction regime between spherical 
(Section \ref{sec:spheres}) and cylindrical (Section \ref{sec:rods}) macroions.  
We  focus on analytical results only based on two exact asymptotic 
theories, namely, the mean-field Poisson-Boltzmann theory and the strong-coupling (SC) 
theory.  The crossover regime between these two limits are considered 
using recent numerical simulations, which are useful
to determine the regime of applicability of both asymptotic theories, 
and also to examine the validity of their predictions. 

We restrict our discussion to a primitive model in which particles only interact
with Coulombic forces and in some cases, also with short-range excluded-volume repulsions.
The inhomogeneous charge distribution of macroions and also ion-surface adsorption effects 
\cite{Belloni95} are neglected. We do not consider coions and 
take the solvent effects into account only through the continuum dielectric constant of medium 
$\varepsilon$ (for water at room temperature $\varepsilon\approx 80$),  which 
is mainly assumed  to be uniform and equal everywhere in space. 
The role of the dielectric jump at charged boundaries will be discussed briefly 
in Section \ref{sec:dielec} for the case of two charged walls.

\section{Length scales in charged systems: from mean-field to strong-coupling regime}
\label{sec:scales}

To distinguish the regimes of parameters where attractive or repulsive forces may
arise between macroions, one needs first to study 
the length scales that appear in a classical charged system.
Let us consider a system of macroions with uniform surface charge density of 
$-\sigma_{\textrm{s}}$ (in units of the elementary charge $e$) 
and neutralizing counterions of charge valency $q$ at 
temperature $T$. (Hereafter, we conventionally assume that macroions are negatively charged and counterions
are positively charged, thus $\sigma_{\textrm{s}}$ and $q$ are both positive by definition.)

A characteristic length scale in such a system is set by comparing 
the thermal energy scale, $k_{\textrm{B}}T$, 
with the Coulombic interaction energy between counterions, 
$V(r)=q^2e^2/(4\pi \varepsilon \varepsilon_0 r)$, where $r$ is the distance between two
given counterions. The ratio between these two quantities may be written 
as $V/(k_{\textrm{B}}T)=q^2\ell_{\textrm{B}}/r$, where 
\begin{equation}
  \ell_{\textrm{B}}=\frac{e^2}{4\pi \varepsilon \varepsilon_0 k_{\textrm{B}}T}
\label{eq:Bj}
\end{equation}
is the so-called Bjerrum length, which
measures the distance at which two elementary charges interact with thermal energy
$k_{\textrm{B}}T$ (in water and at room temperature $\ell_{\textrm{B}}\approx 7.1$\AA). 
Thus the rescaled Bjerrum length 
\begin{equation}
  \tilde \ell_{\textrm{B}}=q^2\ell_{\textrm{B}}
\end{equation}
may be taken as the relevant length scale to characterize the strength of 
mutual counterionic repulsions against thermal fluctuations in the system. 

\begin{figure}[t]
\includegraphics[angle=0,scale=0.5]{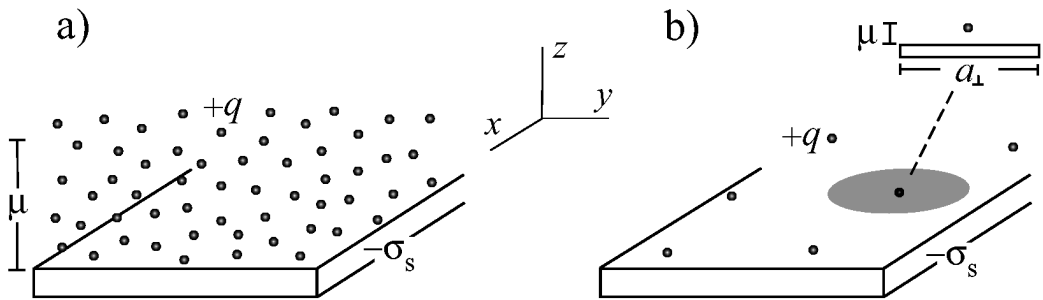}
\caption{a) For small coupling parameter, $\Xi\ll 1$, counterions form a 
diffuse three-dimensional layer. b) For large coupling parameter, 
$\Xi\gg 1$, the counterionic layer is essentially two dimensional since
the typical lateral distance between counterions, $a_\bot$, becomes much
larger than the Gouy-Chapman length, $\mu$. In this regime, counterions 
are strongly correlated and surrounded by a correlation hole of size 
$\sim a_\bot\gg \mu$.}
\label{fig:onewall_layers}
\end{figure}

Other length scales are set by considering the charge distribution and
the specific geometry of macroions. For simplicity, let us concentrate here on a {\em planar}
system composed of a  planar charged wall of  infinitely-large extension and neutralizing counterions
confined to one half-space (Figure \ref{fig:onewall_layers}). 
This model is physically relevant for charged membranes and also for
macroions with large radii of curvature, which behave like plates 
at small distances from their surface.
 
In this system, an additional length scale may be obtained by comparing 
the thermal energy $k_{\textrm{B}}T$ with the energy scale of the counterion-wall attraction, 
$U(z)=q\sigma_{\textrm{s}} e^2 z/(2 \varepsilon \varepsilon_0)$, where $z$ is the vertical distance
from the wall. Hence we have the ratio 
$U/(k_{\textrm{B}}T)=z/\mu$, where 
\begin{equation}
  \mu = \frac{1}{2\pi q \ell_{\textrm{B}} \sigma_{\textrm{s}}}
\label{eq:mu_wall}
\end{equation}
gives the so-called Gouy-Chapman length, which measures
the distance at which the thermal energy equals the counterion-wall 
interaction energy. The Gouy-Chapman length also gives a measure of the 
thickness of the counterionic layer at a charged wall, $\langle z\rangle\sim \mu$, as we shall see later.

In principle, one may tune the system parameters in such a way that either of the two length scales
$\tilde \ell_{\textrm{B}}$ and $\mu$ become arbitrarily large or small. In fact, only the 
dimensionless ratio between these two quantities is relevant and uniquely 
describes the physical regimes of the system \cite{Note1}, namely,
\begin{equation}
  \Xi=\frac{\tilde \ell_{\textrm{B}}}{\mu}=2\pi q^3 \ell^2_{\textrm{B}} \sigma_{\textrm{s}},
\label{eq:Xi}
\end{equation}
which is known as the {\em electrostatic coupling parameter}.
For small coupling parameter $\Xi\ll 1$, 
the counterion-wall system has a relatively large Gouy-Chapman length (or
small Bjerrum length), which indicates 
a loosely bound counterion cloud at the charged wall 
(Figure \ref{fig:onewall_layers}a). 
For large coupling parameter $\Xi\gg 1$, 
the Gouy-Chapman length is relatively small 
(or the Bjerrum length is large) indicating that
counterions are strongly attracted toward the charged wall
(Figure \ref{fig:onewall_layers}b). 

Further insight into the structure of the counterionic layer may be obtained  
by considering the typical distance between counterions at the surface. 
For counterions residing near the charged surface, the local electroneutrality 
condition implies a typical lateral separation of 
\begin{equation}
   a_\bot\sim \sqrt{\frac{q}{\sigma_{\textrm{s}}}},
\label{eq:abot}
\end{equation}
since each  counterion neutralizes the charge of an area given by 
$a_\bot^2\sim q/\sigma_{\textrm{s}}$ (up to a
geometrical prefactor of the order one).
Comparing this length scale with the Gouy-Chapman length, we have 
\begin{equation}
  \frac{a_\bot}{\mu}\sim \sqrt{\Xi}.
\end{equation}
Hence in the strong-coupling regime, $\Xi\gg 1$, counterions essentially
form a quasi two-dimensional layer 
as their lateral separation at surface becomes much larger
than the Gouy-Chapman length $a_\bot\gg \mu$ (Figure \ref{fig:onewall_layers}b). 
On the other hand, the structure of such a layer is dominated by mutual 
repulsions between counterions, which freeze out  lateral degrees of freedom.
Hence counterions become laterally correlated and surrounded by a 
large {\em correlation hole} of size $a_\bot$ 
from which neighboring counterions are statistically depleted \cite{Rouzina96,Shklovs99a,Shklovs02} (see 
Section \ref{subsec:onewall_inter} for a more detailed analysis
\cite{AndreEPJE}). 
This indicates a trend toward crystallization in the ionic structure 
for increasing coupling parameter, 
which can be corroborated by considering the effective
plasma parameter relevant for this situation 
\begin{equation}
  \Gamma=\frac{\tilde \ell_{\textrm{B}}}{a_\bot}\sim \Xi^{1/2}.
\end{equation}
The parameter $\Gamma$ gives a measure of mutual Coulombic repulsions between counterions 
at a neutralizing  surface (the 2D one-component plasma) 
\cite{Rouzina96,Shklovs99a,Shklovs02}. 
For increasing $\Xi$, $\Gamma$ increases while the average counterion-wall
interaction (per $k_{\textrm{B}}T$) remains of the order of unity,
$\langle U\rangle/(k_{\textrm{B}}T)=\langle z\rangle/\mu\sim 1$.
The Wigner crystallization of the 2D one-component plasma is known to occur for
$\Gamma>\Gamma_c\approx 125$ \cite{Baus}, which corresponds to the range of coupling parameters
$\Xi>\Xi_c\approx 3.1\times 10^4$ \cite{AndreEPJE} (see Section \ref{subsec:onewall_inter}).

In the weak-coupling regime ($\Xi\ll 1$), no crystallization is 
expected to occur and the counterion layer has a three-dimensional 
fluid-like structure  ($a_\bot\ll \mu$) \cite{Note2} (Figure \ref{fig:onewall_layers}a).
Thus the two asymptotic regimes of weak coupling ($\Xi\ll 1$) and strong coupling ($\Xi\gg 1$) may be 
distinguished physically by the structure of counterionic layers at charged surfaces. 
In Sections \ref{sec:onewall} and \ref{sec:twowalls},
 we shall briefly review the main results obtained in each of these regimes for the classical example 
of counterions at one and two charged walls (planar double layers). 

But before proceeding  further, it is useful to consider the typical values of the coupling 
parameter in realistic systems. In Table \ref{tab:real_parameters}, we show few 
typical examples of both weakly-coupled and strongly-coupled systems. As already seen from 
Eq. (\ref{eq:Xi}), the coupling strength grows quite rapidly with the counterion valency ($\Xi \sim q^3$), 
which agrees with experimental and numerical evidence indicating highly growing correlation effects 
for increasing counterion valency \cite{Khan,Wennerstroem91,Quirk,Bloom,Kekicheff93,Delsanti,Della,Dubois,Strey98,Tang,Tang03,Angelini03,Guld84,Svensson,Bratko86,Guld86,Wood88,Valleau,Kjellander92,Lyub95,Gron97,Gron98,Wu,AllahyarovPRL,Stevens99,LinsePRL,Linse00,Hribar,Messina00,Allahyarov00,AndreEPL,AndrePRL,AndreEPJE,Deserno03,Naji_epl04,Lee04}. In fact as known from these studies, 
typical coupling strength of $\Xi\sim 10^2$ (or larger) already reflects strong-coupling regime 
and a value of $\Xi\sim 1$ (or smaller) typically corresponds to the weak-coupling regime. 

\begin{table*}[t]
\begin{center}
\begin{tabular*}{12.cm}{l| c c c| c| c c c c c} 
\hline\hline
Charged object & $\sigma_{\textrm{s}}$ ($e$/nm$^2$) && $R$(\AA) & $q$ & $\mu$(\AA) & & $\Xi$ & &
$\xi$ 
\\ \hline
charged membranes & $\sim 1$ && -- & 1 & 2.2 & & 3.1 && -- \\
& &  && 2 & 1.1 && 24.8& & -- \\
& & && 3 & 0.7 && 83.7 & &-- \\
\hline
DNA  & 0.9 && 10 & 1 (Na$^+$) & 2.4 && 2.8 && 4.1   \\
&  &&& 2 (Mn$^{2+}$) & 1.2 && 22.4 & & 8.2   \\
& && & 3 (spermidine) & 0.8 && 75.6  && 12.3   \\
& & && 4 (spermine) & 0.6 && 179 && 16.4   \\
\hline
highly charged colloids & $\sim 1$ && 20 & 3  & 0.7 && 85 && 28   \\
(surfactant micelles) & & && & &&& &\\
\hline
weakly charged colloids &  $\sim 0.1$&& $\sim 10^3$ &1& $\sim 2$  & &$\sim 0.1$ & &  $\sim 5\times 10^2$     \\
(polystyrene particles) & & && & &&& &\\
\hline
\end{tabular*}
\end{center}
\caption{\label{tab:real_parameters} Typical values of physical parameters for realistic charged systems:
$\sigma_{\textrm{s}}$ and $R$ denote the surface charge density and the radius of curvature of charged
objects, $q$ is the charge valency of counterions,  and 
$\mu$, $\Xi$ and $\xi$ are the Gouy-Chapman length, $ \mu = 1/(2\pi q \ell_{\textrm{B}} \sigma_{\textrm{s}})$
(Eq. (\ref{eq:mu_wall})), the coupling parameter, $\Xi=q^2 \ell_{\textrm{B}}/\mu$ (Eq. (\ref{eq:Xi})), and the
Manning parameter, $\xi=R/\mu$ (Eq. (\ref{eq:xi})), respectively. (The role of curvature and Manning parameter 
for cylindrical and spherical macroions is discussed in Sections \ref{sec:curve}-\ref{sec:rods}.) 
The Bjerrum length is taken here as $\ell_{\textrm{B}}\approx 7.1$\AA\, corresponding to an 
aqueous medium of dielectric
constant $\varepsilon=80$ at room temperature.}
\vspace*{5mm}
\end{table*}

\section{Counterions at A charged wall}
\label{sec:onewall}

\subsection{Weak-coupling (or Mean-field) regime $\Xi\ll 1$}
\label{subsec:onewall_PB}

For small coupling strength, one may employ the mean-field approximation to describe
the counterionic layer since each counterion 
interacts with a diffuse cloud of other counterions. 
The mean-field approximation systematically neglects correlations among
counterions and is formally valid in the strict limit of $\Xi\rightarrow 0$ \cite{Netz-orland}.

The mean-field theory is governed by the so-called Poisson-Boltzmann (PB) equation
\cite{VO,Israelachvili,Netz-orland}
\begin{equation}
  \nabla^2 \psi({\mathbf x})=
         -\frac{q e \rho_0}{\varepsilon \varepsilon_0} \exp(-q e \psi/k_{\textrm{B}} T),
\label{eq:PBeq}
\end{equation}
which is to be solved for the mean electrostatic potential in space, $\psi$, 
using proper boundary conditions at macroion surfaces. 
The corresponding density profile of counterions, 
$\rho_{\textrm{PB}}({\mathbf x})$, 
follows from the solution of the PB equation (\ref{eq:PBeq}) and using the relation
$\rho_{\textrm{PB}}({\mathbf x})=\rho_0 \exp(-q e \psi/k_{\textrm{B}} T)$, where
$\rho_0$ is a normalization prefactor.

For the system of 
point-like counterions at a single uniformly-charged wall (in the absence of salt), 
the PB theory predicts an algebraically-decaying density profile of the form  \cite{VO,Israelachvili}
\begin{equation}
  \frac{\rho_{\textrm{PB}}(z)}{2\pi\ell_{\textrm{B}}\sigma_{\textrm{s}}^2}=\frac{1}{(z/\mu+1)^2},
\label{eq:onewall_PBdens}
\end{equation}
where $z$ is the distance from the wall. The density of counterions at contact is obtained
as $\rho(z=0)= 2\pi\ell_{\textrm{B}}\sigma_{\textrm{s}}^2$, which is an exact result within 
the present model and valid beyond the mean-field level \cite{contact_value}. 
As seen the PB theory predicts an extended counterionic density profile  (with diverging 
moments) in agreement with the qualitative considerations in Section \ref{sec:scales} 
for a weakly-coupled system. Nonetheless, 
the PB density profile (\ref{eq:onewall_PBdens}) is normalizable to the total
number of counterions reflecting the fact a charged wall binds all its counterions. 
Note also that the Gouy-Chapman length, $\mu$, in this case equals 
the height of a layer at the wall which contains half of the counterions, and thus 
may be associated  with the typical layer thickness at low couplings. 

\subsection{Strong-coupling regime $\Xi\gg 1$}
\label{subsec:onewall_SC}

In this regime, the liquid-like ordering (or crystallization at sufficiently large couplings) 
of counterions renders the mean-field theory an invalid description of the system. 
Yet one can obtain a simple analytical description for the counterionic layer as follows
\cite{Rouzina96,Shklovs99a,Shklovs02,Netz01} .  

Since for $\Xi\gg 1$ counterions become highly separated from each other 
in a quasi-2D layer at the charged wall  (Figure \ref{fig:onewall_layers}b), 
one may consider the system as a collection of laterally frozen {\em correlation cells}, each consisting 
of a single counterion interacting with an area of the wall of size 
$\sim a_\bot$ (Eq. (\ref{eq:abot})). Since in this regime $a_\bot\gg \mu$, the dominant 
contribution to the density 
profile of counterions at the wall is obtained 
by considering only the vertical degree of freedom, $z$, through which 
single counterions are coupled to the wall with the interaction potential
$U/(k_{\textrm{B}}T)\approx z/\mu$. Hence using the Boltzmann weight, one 
has  the following density profile
\begin{equation}
  \rho_{\textrm{SC}}(z)=\rho_0\exp(-z/\mu).
\label{eq:onewall_SCdens}
\end{equation}
The prefactor in the above expression (the contact density) may be fixed from the normalization
condition for the density profile and the global electroneutrality of the system as 
$\rho_0= 2\pi\ell_{\textrm{B}}\sigma_{\textrm{s}}^2$.
The strong-coupling density, $\rho_{\textrm{SC}}(z)$, 
drops quite rapidly as one moves away
from the charged wall. The average distance of counterions is obtained to be equal
to the Gouy-Chapman length, $\langle z\rangle_{\textrm{SC}} =\mu$.

The above density profile, which essentially follows from single-particle contributions,
was obtained by Shklovskii \cite{Shklovs99a,Shklovs02} using a Wigner-crystal model
for large Coulombic coupling. The asymptotic analysis of Ref. \cite{Netz01}
showed that the partition function of the system for $\Xi\rightarrow \infty$ adopts a series
expansion in powers of $1/\Xi$, the leading term of which is given only by single-particle
contributions. The multi-particle contributions enter in higher-order terms (in the form
of a virial expansion). The leading term defines the {\em asymptotic} strong-coupling
(SC) theory, which for the counterion-wall system gives exactly the density profile (\ref{eq:onewall_SCdens}) . 

\begin{figure}[t]
\includegraphics[angle=0,scale=0.4]{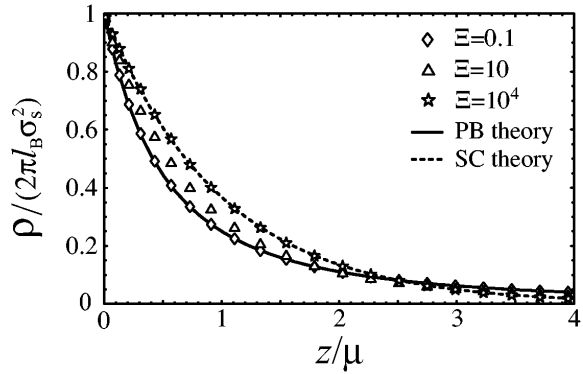}
\caption{Density profile of counterions at a charged wall as a function of the 
distance from the wall, $z$. The symbols are the data from Monte-Carlo 
simulations for $\Xi=0.1$ (open diamonds), $\Xi=10$ (open triangles) 
and $\Xi=10^4$ (open stars). The solid curve and the dashed curve
represent the predictions of the mean-field (PB) theory, 
Eq. (\ref{eq:onewall_PBdens}), and the strong-coupling theory,
Eq. (\ref{eq:onewall_SCdens}), respectively. 
The density profile is rescaled by its values at contact 
$\rho(z=0)= 2\pi\ell_{\textrm{B}}\sigma_{\textrm{s}}^2$ and the distance
from the wall is shown in units of the Gouy-Chapman length, $\mu$.}
\label{fig:onewall_dens}
\end{figure}

\subsection{Intermediate-coupling regime}
\label{subsec:onewall_inter}

In realistic systems, the coupling parameter is always finite (see Table \ref{tab:real_parameters}). 
Therefore, it is important to examine whether and how the preceding analytical
results for the two limits of strong ($\Xi\rightarrow \infty$) and 
weak ($\Xi\rightarrow 0$) coupling may be applied to such systems. 
A useful approach to investigate the regime of intermediate couplings is 
to employ numerical simulation methods. Other methods include systematic improvement
of both mean-field and strong-coupling theories \cite{Netz01} that will be considered only 
for the interaction between two walls in Section \ref{sec:twowalls}. 

The density profile of counterions at an infinitely large 
charged wall has been calculated for various coupling
parameters using Monte-Carlo simulations in Ref. \cite{AndreEPL}.  (The simulation model 
is similar to what we described in Section \ref{sec:scales} (Figure \ref{fig:onewall_layers}),
where counterions are taken as point-like particles confined to one half-space and periodic
boundary conditions are used in lateral directions \cite{AndreEPJE}.) In Figure \ref{fig:onewall_dens}, 
we show the simulated density profile of counterions (symbols) for coupling
parameters $\Xi=0.1, 10$ and $10^4$ along with  the analytical predictions of the mean-field PB theory
(solid curve, Eq. (\ref{eq:onewall_PBdens})) and the strong-coupling theory
 (dashed curve, Eq. (\ref{eq:onewall_SCdens})). 
 Note that in the Figure the distance from the wall is rescaled
with the Gouy-Chapman length; hence, the PB and SC profiles appear to have a similar 
decay length of about unity in rescaled units.
As seen, the data crossover from the mean-field (PB) prediction to the SC result  by
increasing $\Xi$, quantitatively confirming the validity of both theories
at small couplings (about $\Xi=0.1$) and large couplings (about
$\Xi=10^4$) respectively. 

Further analysis of the simulated density profile \cite{AndreEPJE} reveals a 
distance-dependent crossover at intermediate coupling parameters: 
while the large distance behavior of the system
is described well by the mean-field theory, at sufficiently small distances from the 
wall, the system roughly follows the strong-coupling prediction at intermediate $\Xi$. 
This is in fact a quite general property that we shall investigate in more detail
for the interaction between two charged walls in Section \ref{sec:twowalls}.  
The crossover in the single-wall system can be understood qualitatively by noting that
the asymptotic strong-coupling density profile, Eq. (\ref{eq:onewall_SCdens}), 
remains valid  only within distances smaller 
than the correlation hole size $z<a_\bot$ \cite{AndreEPJE,Shklovs99a,Netz01,Burak04}.
Thus at {\em finite} coupling parameters, the strong-coupling regime may be characterized by
\begin{equation}
  \frac{z}{\mu}<\Xi^{1/2},
\end{equation}
where we have used Eq. (\ref{eq:abot}) .  
For distances, $z$, from the wall which are 
larger than $a_\bot$, lateral interactions between  
counterions become relevant and modify the density profile. For  
$z\gg a_\bot$, one intuitively expect that dominant many-body 
effects lead to a mean-field-like behavior as shown in previous studies 
\cite{AndreEPJE,Netz01}.  In fact, the mean-field results turn out to be valid
for distances $z/\mu>\Xi$ (up to some logarithmic 
corrections) \cite{AndreEPJE,Netz01}.

\begin{figure}[t]
\includegraphics[angle=0,scale=0.6]{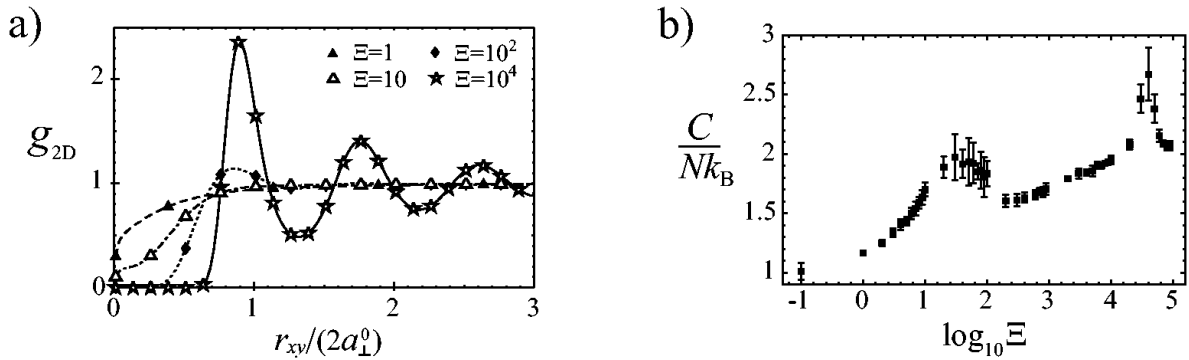}
\caption{a) The two-dimensional pair distribution function of counterions at a 
charged planar wall plotted as a function of the lateral distance between 
counterions (the distribution function is obtained by averaging over the 
height $z$). The symbols are the data from Monte-Carlo 
simulations for $\Xi=1$ (filled triangles), $\Xi=10$ (open triangles),
$\Xi=100$ (filled diamonds) and $\Xi=10^4$ (open stars). The lateral 
distance is rescaled by the length scale 
$2a_\bot^0=2\sqrt{q/(\pi\sigma_{\textrm{s}})}$, which gives a measure of 
lateral separation between counterions (see the text).\\
b) The simulated excess heat capacity of the system of counterions at a 
charged wall (per number of counterions, $N$) plotted as a function of the coupling 
parameter, $\Xi$. (The number of counterions in this case is $N=100$ in a square simulation box, which
is periodically replicated in lateral directions.) The broad hump at intermediate couplings ($10<\Xi<100$) 
reflects the structural change in the counterionic layer due to increasing 
correlations between counterions. At large coupling strength 
($\Xi\approx 3.1\times 10^4$), the counterionic layer undergoes a 
crystallization process indicated by a pronounced peak in the heat capacity. }
\label{fig:onewall_pdfC}
\end{figure}

An interesting problem is to examine the development of correlations between counterions
(including the formation and size of the correlation hole) as the coupling parameter increases.
This goal may be achieved by considering the two-dimensional 
pair distribution function of counterions, which is defined through
\begin{equation}
  g_{\textrm{2D}}=\frac{A}{N^2}\bigg \langle \sum_{\langle ij\rangle} 
           \delta({\mathbf r}_{xy}-{\mathbf r}_{xy, i}+{\mathbf r}_{xy, j})\bigg\rangle,
\end{equation}
where the sum runs over pairs of particles, ${\mathbf r}_{xy}$ is a 2D vector and 
${\mathbf r}_{xy, i}$ is the lateral position of particle $i$ in the 
$xy$-plane (see Figure \ref{fig:onewall_layers}). 
Physically $g_{\textrm{2D}}$ gives the ratio between the probability
of finding two counterions at distance $r_{xy}=|{\mathbf r}_{xy}|$ from each other 
 and the expected probability for a homogeneous 2D fluid with the same 
bulk density. The Monte-Carlo results for this quantity are shown in Figure 
\ref{fig:onewall_pdfC}a \cite{AndreEPJE}. It shows that for small coupling parameter ($\Xi=1$, filled triangles), 
there is only a very short-range depletion zone at small separations between counterions. 
But a pronounced correlation hole is created at the regime
of coupling parameters $10<\Xi<100$,  where the distribution 
function vanishes over a finite range at small inter-particle separations. 
For larger coupling strengths, the correlation hole becomes more pronounced and is followed
by an oscillatory behavior in the pair distribution 
function ($\Xi=10^4$, open stars).  This indicates a liquid-like order in 
the counterionic structure in agreement with qualitative considerations in Section \ref{sec:scales}.
Note that the distance coordinate in Figure \ref{fig:onewall_pdfC}a is rescaled with 
$2a_\bot^0$, where $a_\bot^0=\sqrt{q/(\pi \sigma_{\textrm{s}})}$ is obtained 
by assuming that the area of the wall neutralized by a counterion has a circular
form (Figure \ref{fig:onewall_layers}b). The location of the first peak of 
$g_{\textrm{2D}}$ for  $\Xi=10^4$ appears at a distance of $r_{xy}/(2a_\bot^0)\approx 0.9$, 
which roughly gives the typical lateral distance between counterions. In a perfect hexagonal
crystal, the peak is expected to occur at $r_{xy}/(2a_\bot^0)=\sqrt{\pi/(2\sqrt{3})}\approx 0.95$, and in a
perfect square crystal at $r_{xy}/(2a_\bot^0)=\sqrt{\pi}/2\approx 0.89$. 
The crystallization is in fact found at even larger coupling parameters \cite{Andre_thesis}, which 
may be seen also from the behavior of the heat capacity for increasing $\Xi$.

In Figure \ref{fig:onewall_pdfC}b, the simulated excess heat capacity of the counterion-wall
system (obtained by omitting the trivial kinetic energy contribution $3k_{\textrm{B}}/2$)
is shown for various coupling parameters. The crystallization of counterions
at the wall is reflected by a pronounced peak at large coupling parameters
about  $\Xi_c\approx 31000$, which roughly agrees with
the Winger-crystallization threshold of a 2D one-component plasma \cite{Baus}. The
characteristic properties of the crystallization transition in the counterion-wall system 
are yet to be specified, which requires a detailed finite-size scaling analysis in the vicinity
of the transition point. Another interesting behavior is observed in Figure \ref{fig:onewall_pdfC}b
at the range of coupling parameters  $10<\Xi<100$, where the heat capacity exhibits a broad hump. 
This hump does not represent a phase transition \cite{AndreEPJE}, but it is associated with 
the onset of the correlation hole around counterions and the structural change in the counterionic 
layer from being a three-dimensional layer at low couplings to a quasi-2D layer at large couplings. 
In the region between the hump and 
the crystallization peak  (for $200<\Xi<10^4$),
the heat capacity is found to increase
almost logarithmically with $\Xi$. The reason for this behavior is at present not clear.

\section{Repulsive and attractive like-charge interactions: planar geometry}
\label{sec:twowalls}

\begin{figure}[t]
\includegraphics[angle=0,scale=0.5]{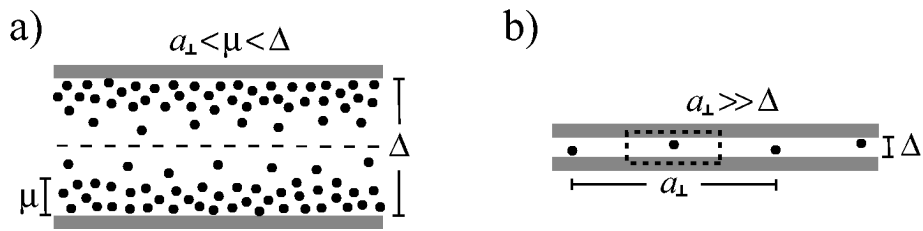}
\caption{Schematic representation of the asymptotic interaction regimes of a) mean field and 
b) strong coupling for two like-charged walls. The mean-field regime is 
obtained at large separations between the walls (compared to other length 
scales) and is dominated by the repulsive osmotic pressure of counterions.
For small wall separation (compared with the typical counterion 
spacing), the walls attract each other since
counterions are isolated in correlation cells of large 
lateral extension $\sim a_\bot/2\gg \Delta$ (shown by a dotted loop)
and mediate a dominant single-particle attraction between the walls.}
\label{fig:twowalls_PBSC}
\end{figure}

Now let us consider the interaction between macroions in the two regimes
of mean field  and strong coupling. In this Section, we focus on the
the planar system of two parallel charged walls of uniform surface 
charge density $-\sigma_{\textrm{s}}$ at separation $\Delta$ from each, where 
$q$-valent counterions fill only the space between the walls (the dielectric constant is also
assumed to be uniform in space)--see Figure \ref{fig:twowalls_PBSC}. 
This model is relevant for interaction between two charged membranes or
two macroions of large radii of curvature.

In this system, an extra length scale is set by the wall separation, $\Delta$. 
Two limiting regimes of  repulsion and attraction may be distinguished qualitatively 
by comparing $\Delta$ with other length scales of the system as follows.

\subsection{Mean-field regime: repulsion}
\label{subsec:twowalls_PB}

First consider the limit where the wall separation, $\Delta$, is {\em large} 
compared with all other length scales in the system and also that 
the system is weakly coupled, $\Xi\ll 1$ (see Figure \ref{fig:twowalls_PBSC}a).
In this case, counterions form a diffuse layer at each wall,
but due to large wall separation, the system is approximately decoupled into two 
nearly neutral sub-systems, each consisting of a charged wall and its 
counterionic cloud.
The effective pressure acting between the walls is dominated 
by the osmotic pressure of counterions across the mid-plane, since 
the overall electrostatic field at the mid-plane is zero due to the charge 
neutrality of each sub-system. This osmotic pressure is positive
and therefore gives an effective {\em repulsion} between the walls. 

The mid-plane osmotic pressure is proportional to the local density of counterions,
$\rho_{\textrm{mid}}$, following the ideal-gas equation $P=\rho_{\textrm{mid}} k_{\textrm{B}}T$, where
 $\rho_{\textrm{mid}}$ drops roughly 
with the inverse square of the wall separation for large $\Delta$
as it follows from Eq. (\ref{eq:onewall_PBdens}). This yields the scaling form 
of the repulsive pressure between the walls as  $P(\Delta)\sim \Delta^{-2}$.
The formal derivation of the pressure based on the PB equation supports the above result
for large separation. The full PB solution for arbitrary $\Delta$ 
follows as \cite{VO,Israelachvili,Netz01}
\begin{equation}
    \frac{\beta P_{\textrm{PB}}(\Delta)}{2\pi \ell_{\textrm{B}} \sigma_{\textrm{s}}^2} =\Lambda
\label{eq:P_PBfull}
\end{equation} 
(with $\beta=1/k_{\textrm{B}}T$), where $\Lambda$ is determined from the transcendental equation 
$\Lambda^{1/2}\tan[\Lambda^{1/2}(\Delta/2\mu)]=1$. For large $\Delta/\mu\gg 1$, 
the PB solution yields
\begin{equation}
  \frac{\beta P_{\textrm{PB}}(\Delta)}{2\pi \ell_{\textrm{B}} \sigma_{\textrm{s}}^2}
          \approx \left(\frac{\pi\mu}{\Delta}\right)^2,
\label{eq:P_PB}
\end{equation} 
which is expectedly independent of the surface charge density of the walls.

\subsection{Strong-coupling regime: attraction}
\label{subsec:twowalls_SC}

Now we consider a different asymptotic regime in which the system is 
strongly coupled, $\Xi\gg 1$, and the distance between
the walls is  smaller than the lateral spacing between counterions at
each wall $a_\bot\gg \Delta$ (Figure \ref{fig:twowalls_PBSC}b).
 Since counterions are highly separated from each other, 
the two opposite layers  of counterions tend to form an inter-locking 
pattern at equilibrium, where each counterion
from one layer faces a bare area of the other wall with nearly equal 
(but opposite) charge.
It is evident that this pattern leads to an effective
{\em attractive} force between the walls with a purely 
energetic origin. The 
effective pressure acting between the walls may be estimated
using a simple argument that takes into account the highly 
correlated structure of counterions \cite{Rouzina96,Netz01}. 

In the asymptotic limit considered here ({\em i.e.} for $\Xi\gg 1$
and $a_\bot\gg \Delta$), each counterion is essentially confined and isolated 
in a ``correlation cell'' that consists of a single counterion sandwiched between 
two opposing sections of the walls with lateral size of about $a_\bot/2$ 
(Figure \ref{fig:twowalls_PBSC}b). Since $a_\bot\gg \Delta$, 
the effective pressure between the walls is dominated 
by the contribution coming from each single correlation cell and 
lateral interactions between these cells may be neglected.  

The electrostatic energy of the system per cell is the sum of the 
interactions between the two surfaces with each other 
and with the single counterion, which--using the electroneutrality 
condition and the fact that the wall separation is small--follows as 
$\beta u_e\approx 2\pi\ell_{\textrm{B}}\sigma_{\textrm{s}}^2 \Delta$ per unit area. 
This {\em energetic} contribution gives an attractive pressure of 
$\beta P_e\approx -2\pi\ell_{\textrm{B}}\sigma_{\textrm{s}}^2$ between the walls. 
On the other hand, the entropic contribution due to the counterion confinement 
is of the order $S\sim k_{\textrm{B}}\ln \Delta$ (per cell), which generates 
a repulsive component. 
The total pressure between the walls is then obtained by combining these two
effects and may be written as 
\begin{equation}
  \frac{\beta P_{\textrm{SC}}(\Delta)}{2\pi\ell_{\textrm{B}}\sigma_{\textrm{s}}^2}=
      -1+\frac{2\mu}{\Delta}.
\label{eq:P_SC}
\end{equation}
This expression clearly predicts a closely-packed {\em bound state} between the walls
with equilibrium surface separation, $\Delta_\ast$,
being equal to twice the Gouy-Chapman length, {\em i.e.}
\begin{equation}
  \Delta_\ast=2\mu.
  \label{eq:Dmin_walls}
\end{equation} 
The like-charged walls attract each other for $\Delta>\Delta_\ast$ and repel at smaller distances.

The energetic attraction in the strong-coupling regime, which results from structural correlations, 
was first obtained by Rouzina and Bloomfield \cite{Rouzina96} and investigated later 
by several workers \cite{Korny,Arenzon99,Shklovs99a,Shklovs99,Shklovs02,Diehl01,Lau01,Netz01,Levin02,Naji_epje04}. 
The rigorous derivation of the expression (\ref{eq:P_SC}) 
for the pressure was given using the asymptotic strong-coupling theory \cite{Netz01}, which shows that
Eq. (\ref{eq:P_SC}) is  the exact result in the limit of $\Xi\rightarrow \infty$.   

Note that in the presence of counterions of finite diameter $\sigma_{\textrm{ci}}$,
the equilibrium separation between the walls, Eq. (\ref{eq:Dmin_walls}), increases by 
an amount equal to the counterion diameter and reads
\begin{equation}
  \Delta_\ast=\sigma_{\textrm{ci}}+2\mu
\end{equation} 
assuming that counterions interact with a hard-core excluded-volume interaction with the walls.
Clearly, the excluded-volume interaction between counterions themselves is irrelevant
in the strong-coupling regime since counterions are highly separated from each other ($a_\bot\gg \Delta$).

\begin{figure}[t]
\includegraphics[angle=0,scale=0.5]{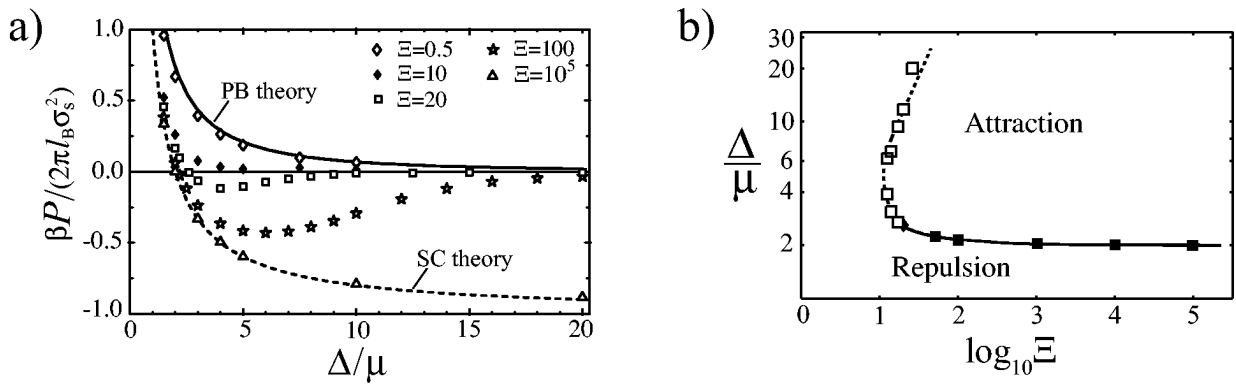}
\caption{a) Effective pressure between two like-charged walls 
as a function of their distance, $\Delta$. The symbols are MC simulation data for 
$\Xi=0.5$ (open diamonds), $\Xi=10$ (filled diamonds), $\Xi=20$ (open 
squares), $\Xi=100$ (open stars) and $\Xi=10^5$ (open triangles). 
The solid curve is the prediction of the mean-field PB theory, 
Eq. (\ref{eq:P_PBfull}), and the dashed curve is the strong-coupling prediction,
Eq. (\ref{eq:P_SC}). The pressure and the wall separation are shown in 
rescaled units as indicated on the graph. \\
b) Attraction and repulsion regimes shown in terms of the rescaled wall distance, $\Delta/\mu$, and the 
coupling parameter, $\Xi$. Symbols are the simulation results indicating 
the zero-pressure points (connecting lines are guides to the eye). Filled 
symbols show the thermodynamically stable bound state of the two 
walls at small separations. 
Open symbols indicate the meta-stable or unstable states of zero
pressure. 
Attraction sets in for $\Xi>12$ and a first-order phase transition occurs 
at $\Xi\approx 17$.}
\label{fig:twowalls_PPD}
\end{figure}

\subsection{Numerical simulations and the crossover regime}
\label{subsec:twowalls_inter}

In order to examine the preceding asymptotic results in the mean-field and 
strong-coupling regimes, we consider the Monte-Carlo simulations of the 
two-wall system, which enable one to investigate the mechanism of 
like-charge interaction beyond  the above limiting cases. 

Figure \ref {fig:twowalls_PPD}a shows the simulated effective pressure 
acting between two like-charged walls (in the presence of point-like counterions) for 
various coupling parameters \cite{AndrePRL,AndreEPJE}. The pressure becomes negative, and thus indicates attraction
between the walls at small and intermediate separations, when the coupling parameter exceeds
an intermediate threshold (see below). 
The onset of attraction at intermediate couplings ($10<\Xi<100$) agrees with the onset of correlations 
between counterions as discussed in Section \ref{subsec:onewall_inter}. 
For small couplings ($\Xi=0.5$, open diamonds), the data quantitatively
support the mean-field PB prediction (solid curve), Eq. (\ref{eq:P_PBfull}),  and for 
very large couplings ($\Xi=10^5$, open triangles), they agree with the SC 
prediction (dashed curve), Eq. (\ref{eq:P_SC}), quite well.

The behavior of the pressure may be summarized in a phase diagram  as 
shown in Figure \ref {fig:twowalls_PPD}b, which shows the regions of positive (repulsive) and 
negative (attractive) pressure separated by a line of zero pressure. 
The attractive region only appears for coupling parameters $\Xi>12$.
The filled symbols (connected with a solid line) show the stable bound state of the 
two walls, while the open symbols (connected with a dashed line) correspond
to meta-stable or unstable states of the two wall ({\em i.e.} the local minima 
or the maximum of the free energy of the system, where the free energy is obtained from the data 
by integrating the pressure from infinite distance 
to a finite distance, $\Delta$) \cite{AndreEPJE}. As seen, the stable bound state
exhibits an equilibrium  wall separation quite close to the strong-coupling prediction,
$ \Delta_\ast/\mu=2$ (Eq. (\ref{eq:Dmin_walls})),
for moderate to large coupling parameters. 
The thermodynamic behavior of this system has been studied in Ref. \cite{AndreEPJE}, 
which predicts a first-order unbinding transition at $\Xi\approx 17$.

The effective pressure also exhibits a distance-dependent crossover 
at intermediate couplings \cite{AndrePRL,AndreEPJE,Netz01}:
at small wall separations, the data closely follow the SC curve, while for large separations, 
they tend to the PB curve and display a mean-field-like repulsion.
In order to study the crossover behavior analytically, 
one needs to consider the extension of both asymptotic theories of mean field 
($\Xi\rightarrow 0$) and strong coupling ($\Xi\rightarrow \infty$) to finite-coupling situations. 

\subsubsection{loop expansion: sub-leading corrections to the mean-field PB theory}
\label{subsubsec:twowalls_loop}

The mean-field PB theory is obtained from a saddle-point approximation in the
limit of $\Xi\rightarrow 0$ \cite{Netz-orland}. Therefore, one way
to incorporate finite-coupling effects on a systematic level is to calculate 
the higher-order corrections to the saddle-point solution by means of a loop expansion. 
The loop parameter turns out to be the coupling parameter $\Xi$ and the 
effective pressure between the walls may be expanded about the mean-field
solution as 
\begin{equation}
  P(\Delta)=P_{\textrm{PB}}(\Delta)+\Xi P^{(1)}_{\textrm{PB}}(\Delta)+{\mathcal O}(\Xi^2),
\label{eq:P_PBexp}
\end{equation}
where $P_{\textrm{PB}}$ is the PB solution (\ref{eq:P_PBfull}), and $P^{(1)}_{\textrm{PB}}$
is the first-loop or the Gaussian correction term \cite{Attard-etal2,Podgornik89}
\begin{equation}
  \frac{\beta P^{(1)}_{\textrm{PB}}}{2\pi\ell_{\textrm{B}}\sigma_{\textrm{s}}^2} 
           \approx -\left(\frac{\mu}{\Delta}\right)^3
                     \bigg [ \frac{\zeta(3)}{4}+\frac{\pi^3}{4}+ \pi^2\ln (\Delta/\pi \mu)\bigg].
\end{equation}

Clearly, the Gaussian correction term contributes an attractive component, which 
comes from correlations between fluctuations in the counterionic clouds 
at opposite walls. These fluctuations
tend to polarize each other giving rise to attraction in the same way as 
other fluctuation-induced attractive forces (such as dispersion interactions)
are generated \cite{Kardar99}.
It is tempting to argue that the Gaussian correction term turns the net pressure
between the walls into an attractive pressure for large enough $\Xi$.
However, the onset of attraction in fact signals the break-down of the 
loop-expansion scheme as used above, since the correction term 
becomes comparable to the leading PB term \cite{Netz01}.
Therefore, the Gaussian-fluctuations picture remains valid only 
at sufficiently small couplings (or the so-called {\em high-temperature regime}) and 
also for sufficiently large separations $\Delta/\mu \gg 1$ 
(see the discussion in Refs. \cite{Ha01,Lau01,Netz01}).

The regime of validity of the loop expansion (and that of the
mean-field PB theory) at a {\em finite} coupling parameter, $\Xi$, may be estimated
by comparing the sub-leading and the leading terms in Eq. (\ref{eq:P_PBexp}), that gives
\cite{AndreEPJE,Netz01}
\begin{equation}
  \frac{\Delta/\mu}{\ln (\Delta/\mu)}>\Xi.
\end{equation}

\subsubsection{virial expansion: sub-leading corrections to the strong-coupling theory}
\label{subsubsec:twowalls_virial}

In the strong-coupling regime, the finite-coupling corrections may be taken into account 
using a virial-expansion scheme, which is obtained as a series expansion in powers of $1/\Xi$ 
about the asymptotic strong-coupling solution (for $\Xi\rightarrow \infty$) \cite{Netz01}. 
The effective pressure between two like-charged walls 
adopts the following large coupling expansion 
\begin{equation}
  P(\Delta)=P_{\textrm{SC}}(\Delta)+\frac{1}{\Xi}P^{(1)}_{\textrm{SC}}(\Delta)+{\mathcal O}(\Xi^{-2}),
\label{eq:P_virial}
\end{equation}
where $P_{\textrm{SC}}$ is the SC prediction, Eq. (\ref{eq:P_SC}), and $P^{(1)}_{\textrm{SC}}$
is the first correction term \cite{AndrePRL,AndreEPJE,Netz01}
\begin{equation}
  \frac{\beta P^{(1)}_{\textrm{SC}}}{2\pi\ell_{\textrm{B}}\sigma_{\textrm{s}}^2} 
           = \frac{\Delta}{3\mu},
\label{eq:P_SCexp}
\end{equation}
which contributes a repulsive component to the total pressure. This finite-coupling 
correction can be used only at sufficiently large couplings and small
wall separations where the correction term itself is small, {\em i.e.} 
where the $1/\Xi$-expansion scheme remains valid.
One may estimate the regime of validity of this expansion
(and thus the regime of applicability of the SC theory) at a {\em finite}
 coupling parameter from Eqs. (\ref{eq:P_virial}) and  (\ref{eq:P_SCexp}) as \cite{Netz01} 
\begin{equation}
   \left(\frac{\Delta}{\mu}\right)^2 < \Xi.
\label{eq:SC_crit}
\end{equation}
This estimate in fact agrees with our qualitative 
discussion in Section \ref{subsec:twowalls_SC}, which predicts the strong-coupling attraction for
small wall separation compared with the  
lateral distance of counterions, $a_\bot$, that is for 
\begin{equation}
  \Delta<a_\bot,
  \label{eq:RB}
\end{equation}
which in units of the Gouy-Chapman length (and using Eq. (\ref{eq:abot}))
reproduces Eq. (\ref{eq:SC_crit}). Note also that the equilibrium wall separation predicted by the SC
theory, $\Delta_\ast/\mu=2$ (Eq. (\ref{eq:Dmin_walls})), fulfills the above criterion for $\Xi>4$.

Equation (\ref{eq:RB})--or (\ref{eq:SC_crit}) in rescaled units--is also 
known as the {\em Rouzina-Bloomfield criterion} \cite{Rouzina96}, 
which is established as a generic attraction criterion for highly-charged macroions
including charged spheres and cylinders \cite{AllahyarovPRL,LinsePRL,Linse00,Naji_epje04,Naji_epl04}. 

\begin{figure}[t]
\includegraphics[angle=0,scale=0.5]{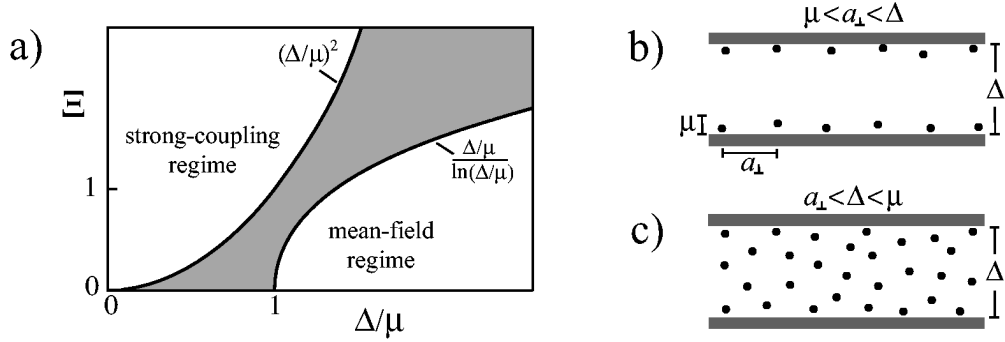}
\caption{a) Regimes of applicability of the asymptotic theories of mean field
Poisson-Boltzmann and strong coupling for the system of two like-charged 
walls and counterions. There is an intermediate regime of rescaled distances 
between the walls, $\Delta/\mu$, and coupling parameters, $\Xi$, where 
finite-coupling effects can not be captured by series expansions around the mean-field
or the strong-coupling solutions. 
Possible physical situations in this regime are schematically shown 
in b) and c)--see the text.}
\label{fig:twowalls_regimes}
\end{figure}

The above discussions may be summarized in a diagram as shown in 
Figure \ref{fig:twowalls_regimes}a  specifying the range of parameters
(coupling parameter and the wall separation) where the strong-coupling or the mean-field 
picture prevails. As seen there appears a gap in the diagram, where neither of the theories can be 
extended to include finite-coupling effects via the series-expansion methods mentioned before.    
The physical situations to which this gap corresponds have been illustrated in 
Figures  \ref{fig:twowalls_regimes}b and  \ref{fig:twowalls_regimes}c (compare these
Figures with Figure  \ref{fig:twowalls_PBSC}).
Figure  \ref{fig:twowalls_regimes}b shows a system in which the Gouy-Chapman length 
is the smallest length scale 
and the wall separation is large such that $\mu<a_\bot<\Delta$. 
In units of the Gouy-Chapman length, we have $1<\Xi<(\Delta/\mu)^2$. 
In this case, the PB approach 
is not valid and in a rough approximation, the two layers are decoupled and each layer
is separately described by the strong-coupling density profile for a single wall.  
Yet a systematic theory for the effective interaction in this regime is missing. 
Figure  \ref{fig:twowalls_regimes}c shows a system in which the Gouy-Chapman length 
is the largest length scale
and $a_\bot<\Delta<\mu$, or in units of the Gouy-Chapman length,
$\Xi<(\Delta/\mu)^2<1$. 
In this case, counterions form a confined gas with local three-dimensional correlations
for finite $\Xi$.
Interestingly in this regime both SC theory and PB theory agree on the leading 
level but again have different corrections \cite{Netz01}.

\section{The role of dielectric jump at charged surfaces}
\label{sec:dielec}

So far we have assumed that the dielectric constant is uniform in space and equal for
both solvent medium (where counterions are present) and charged surfaces. 
However,  charged surfaces (macroions) usually have a  dielectric constant, which is  different
from that of an aqueous solvent; for  bio-soft materials, 
the dielectric constant, $\varepsilon'$, is usually smaller than that of
water  $\varepsilon \approx 80$ ({\em e.g.}  $\varepsilon'\approx 2$ for hydrocarbon). 
This introduces a dielectric jump at charged boundaries, which 
can be treated theoretically using the method of image-charges.

For a counterion of charge valency $q$ at a charged wall of dielectric constant $\varepsilon'$, 
the image-charge is given by 
$q'=q\delta_{\varepsilon}$, where 
\begin{equation}
  \delta_{\varepsilon}=\frac{\varepsilon-\varepsilon'}{\varepsilon+\varepsilon'},
\end{equation}
uniquely represents the dielectric jump. Intuitively, one expects that a dielectric jump
of $\delta_{\varepsilon}>0$ leads to  the depletion of counterions from the vicinity of
the charged wall, since the image-charges have the same sign as 
counterions \cite{Bratko86,Kjellander84,Attard-etal2,Podgornik89,Andre02}. 
The depletion of counterions at a charged wall appears to be weak for small 
coupling parameters (the mean-field regime) and 
becomes significant for increasing coupling parameter (see below) \cite{Andre02,Andre_thesis}. 
The counterionic density profile for $\delta_{\varepsilon}>0$
shows a similar cross-over behavior as in the case of $\delta_{\varepsilon}=0$ (Section \ref{subsec:onewall_inter}) 
when coupling parameter is finite, {\em i.e.} it agrees with the strong-coupling prediction at small distances 
from the wall and follows  the mean-field prediction for large distances (note that 
the strong-coupling theory in this case explicitly includes the image-charges).
We shall not consider the case of a single charged wall here and only focus 
on Monte-Carlo results for the effective interaction 
between two like-charged walls in the presence of a dielectric jump \cite{Swetlana}. 

\begin{figure}[t]
\includegraphics[angle=0,scale=0.38]{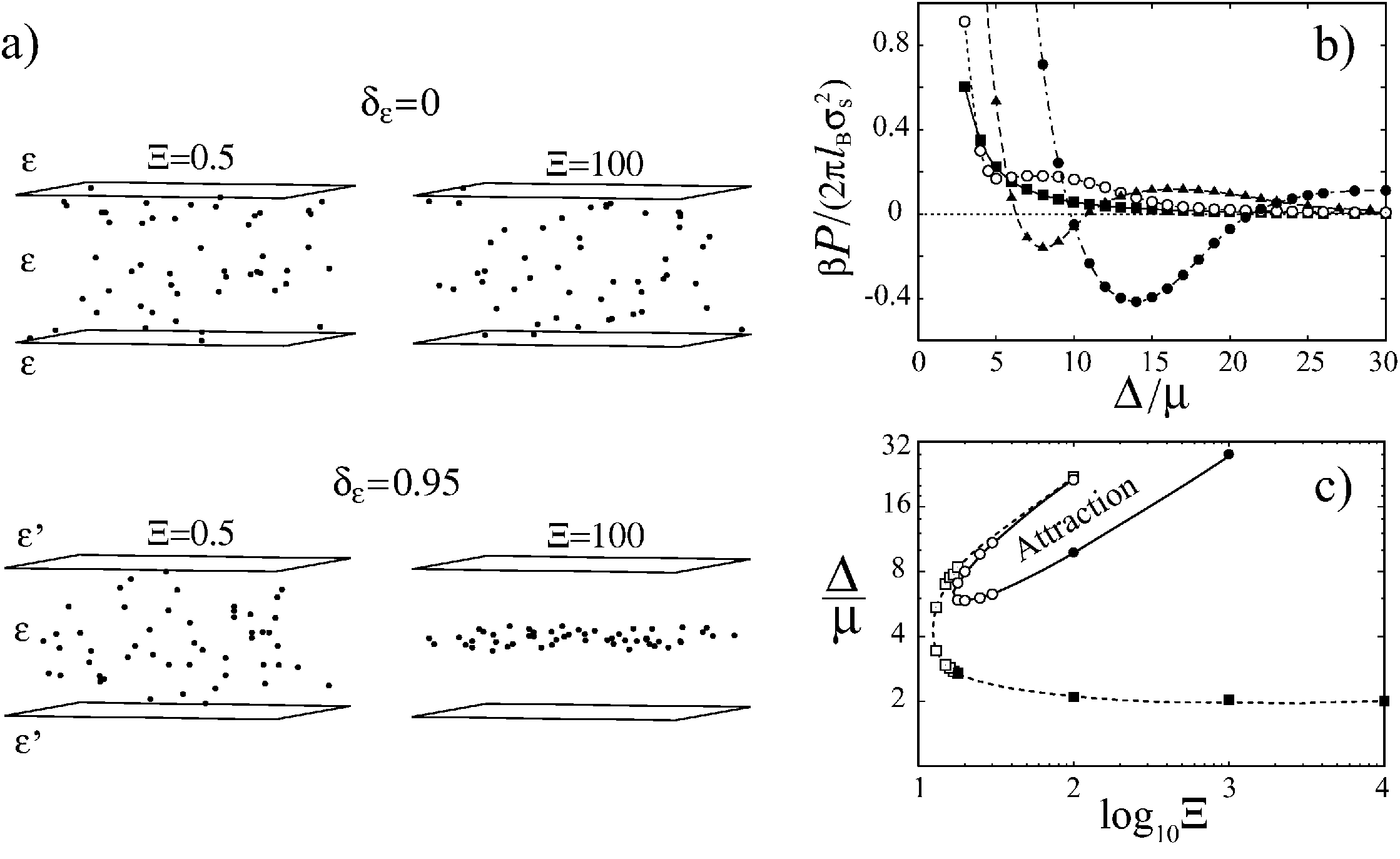}
\caption{a) Snapshots from MC simulations of  two like-charged walls and counterions with ($\delta_{\varepsilon}=0.95$) 
and without ($\delta_{\varepsilon}=0$) a dielectric jump at the walls and for two different coupling parameters as indicated on the
graph. The distance between the walls is $\Delta/\mu=1.0$. The figure shows the main simulation box, which is
replicated periodically in lateral directions in the simulation model. The lateral box size is determined from the global
electroneutrality condition as $L/\mu=\sqrt{\pi \Xi N}$, where $N$ is the number of counterions (here $N=50$). 
For the sake of representation, the extension of the system in lateral directions is rescaled in each case 
with $L/\mu$.\\
b) The rescaled effective pressure between two like-charged walls as a function of the rescaled wall separation, $\Delta/\mu$,
as obtained from simulations in the presence of a dielectric jump of  $\delta_{\varepsilon}=0.95$ at both walls. 
Symbols correspond to different coupling parameters $\Xi=0.5$ (filled squares), 10 (open circles), 30
(filled triangles), 100 (filled circles), and connecting lines are guides to the eye.\\
c) Global behavior of pressure between the walls shown in terms of the rescaled wall distance, $\Delta/\mu$,
and the coupling parameter, $\Xi$. Symbols are the zero pressure points; filled symbols show the stable bound state
of the walls and open symbols show the meta-stable and unstable states \cite{Swetlana}. Circles correspond to the case
with a dielectric jump of $\delta_{\varepsilon}=0.95$ at the walls, 
and squares represent the results for $\delta_{\varepsilon}=0$ (corresponding to Figure \ref{fig:twowalls_PPD}b).}
\label{fig:dielectric}
\end{figure}

Let us consider two like-charged walls of surface charge density $-\sigma_{\textrm{s}}$ and 
point-like neutralizing counterions confined in the space between the walls, which are located at 
a distance of $\Delta$. 
We assume the same dielectric jump of  $\delta_{\varepsilon}>0$ at both walls. 
In this case, one has to account for an infinite number of image-charges for each counterion, 
which leads to quite involved
numerical calculations. We shall therefore proceed with an approximate description by taking into account
only the first-order images (that is one image for each counterion in each wall). 
We show typical snapshots  from Monte-Carlo simulations of the two-wall system with and 
without a dielectric jump in Figure \ref{fig:dielectric}a. The depletion of counterions from the walls
is clearly seen in these snapshots and appears to be stronger for larger coupling strength, since the repulsive
interaction between counterions and their images grows with $\Xi$. 
The effective pressure between the walls is shown in Figure \ref{fig:dielectric}b for a 
dielectric jump of $\delta_{\varepsilon}=0.95$ (corresponding to water-hydrocarbon interface) for various coupling
parameters.  The image-charge interactions lead to a higher pressure at small wall separations as compared to the case with
$\delta_{\varepsilon}=0$ (Figure \ref{fig:twowalls_PPD}a). Also the range of distances at which the two walls attract each other is
pushed to larger wall separations, which is more noticeable for larger coupling strength and 
implies a larger attraction at intermediate distances  (see also Ref. \cite{Bratko86}).  
However for sufficiently large wall separations, the dielectric effects weaken and the pressure becomes repulsive.

The global behavior of the effective pressure is shown in Figure \ref{fig:dielectric}c
 for $\delta_{\varepsilon}=0.95$ (circles) and $\delta_{\varepsilon}=0$ (squares), 
where symbols show the points of zero pressure with filled 
symbols representing the stable bound state of the walls and 
open symbols representing the meta-stable or unstable states (corresponding to the local minimum and 
maximum of the free energy).  
As seen, the onset of attraction is shifted 
to somewhat larger coupling parameters ($\Xi\approx 30$) in the presence of a 
dielectric jump, and also the bound state separation between the walls is larger  
and increases with the coupling parameter.
However, one should be careful in drawing conclusions for the phase behavior
of this system from the present data, since we have only considered the first-order images. 
The full numerical analysis of this system and comparison with extended strong-coupling theory (incorporating
the dielectric jump) will be presented elsewhere \cite{Swetlana}.

\section{The role of curvature: Cylindrical and spherical macroions}
\label{sec:curve}

In the preceding Sections, we considered planar systems, whereas
in realistic situations, charged surfaces often have an intrinsic curvature.
For simplicity, let us consider here only charged spherical and cylindrical macroions which are
characterized by a single radius of curvature $R$. 
The radius of curvature sets a new length scale, which can introduce new features
in terms of counterionic properties of the system. 

Intuitively, one may expect that when the radius of curvature of macroions is
larger than the Gouy-Chapman length, $R\gg \mu$, the
properties of the system remain {\em qualitatively} close to those
of planar charge walls. The qualitative deviations from 
planar case may thus be expected for small $R/\mu$.
As we shall see later, the geometrical symmetries of macroions, 
{\em e.g.} whether they be cylindrical or spherical, also play a
role as they enforce the boundary conditions to which counterions are 
electrostatically coupled. Still it is useful to consider the dimensionless ratio 
of curvature radius, $R$, and the Gouy-Chapman length, $\mu$, as
\begin{equation}
  \xi=\frac{R}{\mu},
\label{eq:xi}
\end{equation} 
which can characterize some important aspects associated with the curvature. 
For curved surfaces of uniform surface charge density $\sigma_{\textrm{s}}$, 
we adopt the same definition for the Gouy-Chapman length 
as in Eq. (\ref{eq:mu_wall}), {\em i.e.}
\begin{equation}
      \mu=\frac{1}{2\pi q \ell_{\textrm{B}} \sigma_{\textrm{s}}}
          = \left\{
             \begin{array}{ll}
               R/(q\ell_{\textrm{B}} \tau)
              & {{\textrm{charged cylinders}},}\\
              \\
              2R^2/(q\ell_{\textrm{B}} Z)
              & {{\textrm{charged spheres}},}
      \end{array}
        \right. 
\label{eq:mu_curve}
\end{equation}
where $\tau=2\pi \sigma_{\textrm{s}} R$ is the linear charge density (in units of the elementary charge $e$)
in the case of charged 
cylinders, and  $Z=4\pi \sigma_{\textrm{s}} R^2$ is the total charge valency for
charged spheres. It is important to note that in these cases, $\mu$ does not necessarily reflect 
the mean distance of counterions from the surface in contrast to the charged walls.

For charged {\em cylinders}, the parameter $\xi$ is quite well-known and is referred
to as {\em Manning parameter} \cite{Manning69}, which may be written as
\begin{equation}
  \xi=\frac{R}{\mu}=q\ell_{\textrm{B}} \tau,
  \label{eq:xi_rods}
\end{equation} 
where we have used Eqs. (\ref{eq:xi}) and (\ref{eq:mu_curve}).
By analogy we shall refer to the same ratio for charged {\em spheres} as 
Manning parameter, which reads
\begin{equation}
  \xi=\frac{R}{\mu}=\frac{q\ell_{\textrm{B}}Z}{2R}.
\label{eq:xi_sph}
\end{equation}

Note that in a system with counterions of finite diameter, $\sigma_{\textrm{ci}}$, 
the effective Gouy-Chapman length is larger than what 
one obtains for the same system with point-like counterions. 
Specifically, when counterions have a hard-core volume interaction with macroions, one has to use
the hard-core radius of macroions in Eq. (\ref{eq:mu_curve}), that is
\begin{equation}
  R_{hc}=R+\sigma_{\textrm{ci}}/2,
\label{eq:Rhc}
\end{equation}
which is larger than the actual radius, $R$, resulting in 
a reduced surface charge density for a given $Z$ (or $\tau$) in the case of spheres (or cylinders). 
This leads to a larger Gouy-Chapman length, Eq. (\ref{eq:mu_curve}), and a 
smaller coupling parameter, Eq. (\ref{eq:Xi}). 
But the Manning parameter as defined in Eq. (\ref{eq:xi}) 
remains unchanged \cite{Naji_epje04}.

\subsection{Binding-unbinding transition of counterions}
\label{subsec:ci_binding}

The peculiar features emerging in the presence of charged curved surfaces are related to the
behavior of counterion at large distances from the surface. At equilibrium, 
counterions tend to diffuse away from 
macroions in order to maximize the entropy of the system, while at the same time, they are
attracted energetically toward the macroion surfaces.  

For counterions at a charged {\em sphere}, 
the gain in entropy grows with the distance of counterions from
the macroion center, $r$, qualitatively as $\sim \ln r$ for large distances. 
The energetic attraction, on the other hand, behaves like $1/r$, which therefore
is always weaker than the {\em entropic repulsion} experienced by counterions. 
Hence, counterions at a charged sphere tend to unbind completely 
and diffuse to infinity in the absence of confining boundaries. 
Thus the role of confinement becomes important in
keeping counterions in the proximity of charged spheres. The confinement
volume per sphere is related inversely to the concentration of 
spherical macroions in a solution. 
The qualitative considerations given above indicate that 
in the {\em infinite-dilution limit} (where the concentration of spheres tends to zero), 
the equilibrium counterionic density profile vanishes due to the
{\em complete de-condensation} of counterions. 
Note that for a charged {\em wall}, the counterion-wall 
attraction grows linearly with distance, $\sim z$, overcoming the entropic contribution.
The charged wall thus binds all its counterions, which is reflected by the fact that the density 
profile of counterions--though extended to infinity as in the 
mean-field regime--is normalizable to 
the total number of counterions (see Sections \ref{subsec:onewall_PB} and \ref{subsec:onewall_SC}). 
We shall investigate these aspects further in the context of interaction 
between two spheres in Section \ref{sec:spheres}.

The case of charged {\em cylinders} lies between the two cases of charged walls and
spheres in that the
energetic attraction of counterions to the cylinder grows logarithmically with 
the distance from the cylinder axis, $r$, {\em i.e.}
in the same way as the entropic gain increases, $\sim\ln r$. 
The competition between these two effects
can result in a threshold binding-unbinding process in this geometry
when the infinite-dilution limit is reached. In order to determine the threshold, one needs to 
consider the prefactor of both logarithmic contributions. 
The energetic attraction of a counterion to 
an infinitely long cylinder 
is given by $\beta U= 2 q \ell_{\textrm{B}} \tau\ln r=2\xi \ln r$  (per $k_{\textrm{B}}T=\beta^{-1}$). The
entropic gain at large separations may be written as $S/k_{\textrm{B}}=2\ln r$, 
since the cylindrical boundary implies a two-dimensional geometry. 
Comparing the two contributions, a threshold value of $\xi_\ast=1$ is 
obtained: for Manning parameter $\xi>1$, the attraction wins and can lead to 
{\em partial} binding of counterions (with a finite density profile at 
the cylinder), whereas for $\xi<1$, complete de-condensation of counterions is 
expected.
This qualitative picture is actually supported by existing analytical 
results of mean-field \cite{Manning69,Zimm} and strong-coupling \cite{Naji_epje04} theory;
both limiting theories give the same threshold of $\xi_\ast=1$. 
Recent numerical simulations \cite{Naji-Netz-unpub} show that this 
threshold is in fact universal and holds in all ranges of the coupling 
parameter $\Xi$. Moreover, the threshold counterion-condensation process at 
charged cylinders exhibits a set of scaling relations, which are characterized
by universal exponents \cite{Naji-Netz-unpub}.

In the case of two charged cylinders, the threshold process is expected to occur at Manning
parameter $\xi=1/2$ \cite{Manning97} as we shall discuss in Section \ref{sec:rods}.

\subsection{Attraction criteria for interacting spheres and cylinders}
\label{subsec:RB}

The binding-unbinding behavior of counterions can drastically 
affect the effective interaction between macroions in solution, particularly,
at low concentrations of macroions. When counterionic 
clouds at macroions become diluted due to the de-condensation process, 
counterion-mediated interactions 
are weakened and the effective 
interaction between macroions is dominated by their bare Coulombic 
repulsion. As mentioned above, the de-condensation process
may in principle occur in all ranges of the coupling parameter, $\Xi$, 
since it is regulated by 
the Manning parameter, $\xi$, which is independent from the coupling
parameter. 
Thus for curved surfaces--in contrast to planar systems--a large coupling 
parameter ($\Xi\gg 1$) by itself 
does not necessarily indicate the regime of large electrostatic 
correlations, where strong-coupling attraction is expected between like 
charges. 

In order to specify the attraction-dominated  
regime for interacting spheres and cylinders, one can still employ a
criterion similar to the Rouzina-Bloomfield criterion introduced for two charged 
walls (see below). But such a criterion should be 
supplemented by an additional condition on Manning parameter
guaranteeing that a sufficiently large fraction 
of counterions condense in the vicinity of macroions.
It is however difficult to establish this latter condition even for the 
simplest interesting cases of two spheres and two cylinders. Because it 
requires a detailed analysis of the binding-unbinding process 
of counterions in these systems, which is available only in 
the asymptotic cases of mean field ($\Xi\rightarrow 0$) \cite{Manning69,Manning97} and 
strong coupling ($\Xi\rightarrow \infty$) \cite{Naji_epje04}. The mean-field  theory 
is irrelevant for our purpose (as it  does not include correlations), 
but as we shall see in Sections \ref{sec:spheres} and \ref{sec:rods}, 
the strong-coupling theory can be used 
to obtain a quantitative prediction for the range of Manning parameters, 
where attraction may occur between macroions. In the following, we briefly 
mention the Rouzina-Bloomfield 
attraction criteria for spheres and cylinders assuming that
the condition on Manning parameter is fulfilled.

\subsubsection{Like-charged spheres}
\label{subsubsec:RB_sph}

When counterions are highly condensed at spheres, we expect that correlation effects become dominant 
when the typical distance between counterions, $a_\bot$, becomes 
much larger than the Gouy-Chapman length. Using the local electroneutrality condition,
one may estimate $a_\bot$ as 
\begin{equation}
  a_\bot\approx R\sqrt{\frac{4q}{Z}}
\label{eq:abot_sph}
\end{equation}
(up to some numerical prefactor of the order unity), 
which also gives a measure of the correlation hole size around counterions at surface
(Figure \ref{fig:spheres_model}). 
Note that in units of the Gouy-Chapman length, $\mu$, we have 
$\tilde a_\bot=a_\bot/\mu\sim (2\Xi)^{1/2}$, where the coupling parameter for charged
spheres may be written as 
\begin{equation}
  \Xi = \frac{q^3\ell_B^2Z}{2R^2},
\label{eq:Xi_sph}
\end{equation}
using Eqs. (\ref{eq:Xi}) and (\ref{eq:mu_curve}). 
The correlation-induced attraction is expected when the surface-to-surface distance between 
spheres, $\Delta$, becomes smaller than the
counterionic separation at the opposing surfaces, {\em i.e.}   
\begin{equation}
  \Delta < a_\bot.  
\label{eq:crit_sph}
\end{equation} 
In units of the Gouy-Chapman length, the above criterion may be written in terms of the coupling parameter as 
$({\Delta}/{\mu})^2 < \Xi$, which is qualitatively similar to the attraction condition obtained for two charged walls 
in Section \ref{subsubsec:twowalls_virial}.
The above attraction criterion 
was explicitly verified in simulations by Allahyarov {\em et al.} \cite{AllahyarovPRL} 
and Linse {\em et al.} \cite{LinsePRL,Linse00}
on like-charged spheres (Section \ref{subsec:spheres_sim}). 
  
\subsubsection{Like-charged cylinders}
\label{subsubsec:RB_rods}

For charged (parallel) cylinders, the attraction is similarly expected to arise for
\begin{equation}
  \Delta < a_z,
\label{eq:crit_rods}
\end{equation}
where $\Delta$ is the surface-to-surface distance of cylinders and $a_z$ 
is the typical separation between condensed counterions \cite{Naji_epl04}. In the 
case of highly-charged cylinders, counterions tend to accumulate in the 
intervening region between cylinders, where they line up on opposing 
surfaces along the cylinder axis forming a correlated inter-locking pattern 
\cite{Deserno03}.
The typical separation between counterions in this situation, $a_z$, 
may be estimated from the local electroneutrality condition $q=\tau a_z$, giving
\begin{equation}
  a_z=\frac{q}{\tau}, 
\end{equation} 
which agrees with the results obtained in recent simulations \cite{Deserno03}. 
We shall discuss the application of criterion (\ref{eq:crit_rods}) in Section \ref{subsec:rods_sim}
using numerical simulations of the two-cylinder system.

\section{Attraction between like-charged spheres}
\label{sec:spheres}

\subsection{Numerical simulations}
\label{subsec:spheres_sim}

Recently, there have been several simulations  \cite{Gron98,Wu,AllahyarovPRL,LinsePRL,Linse00,Hribar,Messina00}
investigating effective electrostatic attraction between like-charged spheres in the large coupling
regime using multivalent counterions or in some cases, using low dielectric constants 
\cite{AllahyarovPRL} or considering the system at low temperatures \cite{Messina00}. 
Like-charge attraction is reported in all these simulations for moderate to large coupling parameters
(see Table \ref{tab:sim_parameters_sph}).  

The strength of attractive force obtained between spheres is sufficiently large that it can lead to
closely-packed bound states (including large aggregates) between like-charged spheres 
\cite{Gron98,Wu,AllahyarovPRL,LinsePRL,Linse00,Hribar,Messina00}. 
The bound-state corresponds
to an attractive minimum in the potential of mean-force between spheres at 
small surface-to-surface separations \cite{Gron98,Wu,LinsePRL,Linse00}. 
An interesting feature is that the attraction
regime at small separations is separated by a pronounced {\em potential barrier}
from a repulsion regime at large separations (see Fig. 1 in Ref.  \cite{Gron98}).  
On the other hand, the attractive minimum and
the potential barrier are not robust and exhibit a dependence upon the size 
of the confinement volume  \cite{Gron98}: for increasing confinement volume,
the depth of the attractive minimum and at the same time
the height of the potential barrier, decreases leading to a long-ranged repulsion 
between spheres in a sufficiently large confinement as expected.
Though such a dependence on confinement volume appears to be quite weak 
(Section \ref{subsec:sph_threshold}). 

The existence of a potential barrier in the interaction potential of {\em confined} 
spheres can result in meta-stable bound states between two highly-coupled spheres 
 \cite{Messina00} and also indicates 
a first-order phase transition (phase separation between a dilute and an aggregated
phase) in the thermodynamic limit \cite{Gron98,LinsePRL,Linse00}. 

We shall mainly focus on the attraction regime at small separations, where 
the attractive force between spheres maintains a typical {\em equilibrium}
surface-to-surface distance of the order of the counterion diameter 
\cite{Gron98,Wu,LinsePRL,Linse00}. 
This equilibrium (or bound-state)
separation may be obtained from the location of the minimum of the potential of mean force
or equivalently from the pair distribution function of spheres. 

In Table  \ref{tab:sim_parameters_sph}, we show the estimated values for the 
equilibrium surface-to-surface distance between
spheres in these simulations, $\Delta_{\textrm{sim}}$, along with other parameters
such as the coupling parameter, $\Xi$, the Gouy-Chapman length, $\mu$, as well as the estimated
typical separation between counterions, $a_\bot$, from Eq. (\ref{eq:abot_sph}). 
As seen the Gouy-Chapman length is quite small  in these systems ($\mu\sim 1$\AA)
compared with the counterion separation at spheres,
$a_\bot$, resulting in a large coupling parameter. 
The strong-coupling attraction criterion (\ref{eq:crit_sph}) is also fulfilled, since
the equilibrium separation between spheres, $\Delta_{\textrm{sim}}$, is 
found to be smaller than $a_\bot$. 
These observations suggest that the asymptotic strong-coupling theory
is indeed relevant in the regime of parameters considered in these simulations.
In the following, we shall present the strong-coupling predictions and compare them 
with the simulation results. 

\begin{table*}[t]
\begin{center}
\begin{tabular*}{15.5cm}{l c c c c c c|c c c c c|c c} 
\hline\hline
Simulation & $q$ & $Z$ & $\ell_{\textrm{B}}$(\AA) & $R$(\AA) & $\sigma_{\textrm{ci}}$(\AA) & 
$L$(\AA) & $\mu$(\AA) & $\xi$  & $\Xi$ & $\xi_c^{(1)}$  
& $a_\bot$(\AA) & $\Delta_{\textrm{sim}}$(\AA) & $\Delta_\ast$(\AA)
\\ \hline
Gr{\o}nbech-Jensen {\em et al.} \cite{Gron98} & 2 & 10 & 7.01 & 7& 3.3& 
50-200& 1.07 & 8.1 & 26  & 2.8-3.7  & 7.7 & 2.5 & 3.41   \\
Wu {\em et al.} \cite{Wu} & 2 & 20 & 7.14 & 10 & 4 & 100 & 1.01 & 11.9  & 28  &  
3.3 & 7.5 &  4 & 4.10 \\  
Allahyarov {\em et al.} \cite{AllahyarovPRL} &2 & 32 & 112 & 48.9 & 4.4 & 
$\sim 10^2$ & 0.73 & 70.1 & 615   & $\sim 3.3$ &  25.5 & -- & 4.41  \\
Linse {\em et al.}  \cite{LinsePRL,Linse00} & 3 & 60 & 7.15 & 20 & 4 & $\sim 
10^2$ & 0.75 & 29.2 & 85  & $\sim 3.3$ & 9.8 &  4 & 4.05  \\
Hribar  {\em et al.} \cite{Hribar}  & 3 & 12 & 7.15 & 10 & 2 & $\sim 10^2$ & 
0.94 & 11.7 & 68 & $\sim 3.3$  & 11.0 &  -- & 2.09 \\ 
\hline\hline 
\end{tabular*}
\end{center}
\caption{\label{tab:sim_parameters_sph} Parameters from simulations on 
highly-charged spheres: $q$ is the charge valency of counterions with diameter $\sigma_{\textrm{ci}}$, 
$Z$ is the charge valency of spheres with radius $R$, and $\ell_{\textrm{B}}$,
$\mu$, $\xi$ and $\Xi$ are the Bjerrum length, Eq. (\ref{eq:Bj}), Gouy-Chapman length,
Eq. (\ref{eq:mu_curve}), the Manning parameter, Eq. (\ref{eq:xi_sph}), and the
coupling parameter, Eq. (\ref{eq:Xi_sph}), respectively. $L$ is the confinement box size
and $\xi_c^{(1)}$ is the estimated attraction threshold discussed in 
Section \ref{subsec:sph_threshold}. The last two columns show the equilibrium 
surface-to-surface distance obtained
in these simulations, $\Delta_{\textrm{sim}}$ (if explicitly measured), and
the corresponding result from the strong-coupling theory, $\Delta_\ast$. 
Some of the numbers are given up to the order of magnitude, and the 
extracted values of $\Delta$ from simulations
have a typical resolution of about 1\AA. Note also that in
estimating the values of $\Xi$, $\mu$ and $a_\bot$, we account for the finite size of counterions
by assuming that they have a hard-core interaction 
with macroions  \cite{Wu,AllahyarovPRL,LinsePRL,Linse00,Hribar}--see the note
before Section \ref{subsec:ci_binding} and Ref. \cite{Naji_epje04}.  }
\vspace*{5mm}
\end{table*}

\begin{figure}[t]
\includegraphics[angle=0,scale=0.6]{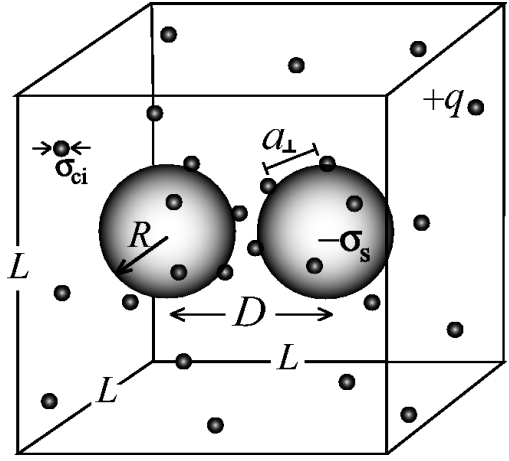}
\caption{Two identical like-charged spheres of radius $R$ are considered at 
center-to-center distance of $D$ in a cubic box of edge size $L$. The bare 
charge of spheres with uniform surface charge density $-\sigma_{\textrm{s}}$ 
(in units of the elementary charge, $e$, and assumed to be negative)
is compensated by the total charge of counterions of charge valency $q$ (and diameter
$\sigma_{\textrm{ci}}$). }
\label{fig:spheres_model}
\end{figure}

\subsection{Effective sphere-sphere interaction: Asymptotic strong-coupling theory}
\label{subsec:spheres_SC}

Let us consider a system of two like-charged spheres 
(Figure \ref{fig:spheres_model}) that are in general confined in 
a cubic box of edge size $L$ inside which the global electroneutrality 
condition is satisfied, {\em i.e.} $N q = 2 Z$, where $N$ is the number of counterions  and  
$Z=4\pi R^2 \sigma_{\textrm{s}}$ is the (absolute value of the) charge valency of each sphere. We assume
that counterions (of diameter $\sigma_{\textrm{ci}}$) 
have also a hard-core excluded-volume interaction with the spheres. 

We are interested in the limit of large couplings. As mentioned in Section \ref{subsec:twowalls_SC}, 
the leading contribution to the free energy for $\Xi\rightarrow \infty$ 
involves only the one-particle contributions. The strong-coupling free energy
 for the two-sphere system
may be written as (up to an irrelevant additive term)  \cite{Naji_epje04}
\begin{equation}
  \frac{\beta {\mathcal F}_{\textrm{SC}}}{N}=\frac{\xi^2}{\tilde D}-
                          \ln  \int_V\,{\textrm{d}}x\,{\textrm{d}}y\,{\textrm{d}}z\,e^{-\beta u(x,y,z)},
\label{eq:SCfree_sph}
\end{equation}
where $\tilde D=D/\mu$ is the center-to-center distance of spheres in units of
the Gouy-Chapman length. 
The first term in Eq. (\ref{eq:SCfree_sph})
is the bare Coulombic repulsion of spheres with Manning parameter $\xi$.
In the second term, $ \beta u$ is the single-particle interaction energy of counterions with 
the spheres, 
\begin{equation} 
          \beta u= -2\xi^2\,
                               (\frac{1}{{\tilde r}_1}+
                                 \frac{1}{{\tilde r}_2}),
\end{equation}    
in which $\tilde r_1=r_1/\mu$ and $\tilde r_2=r_2/\mu$ are the rescaled radial distances from the centers
of the two spheres labeled 1 and 2  (we choose the frame of reference in the middle of the box
such that $r_{1,2}=[(x\pm D/2)^2+y^2+z^2]^{1/2}$). 
The spatial integral in Eq. (\ref{eq:SCfree_sph}) runs over the volume accessible 
for counterions, {\em i.e.} inside the cubic box excluding the two spheres
(note that a shell of thickness $\sigma_{\textrm{ci}}/2$ around each sphere 
is also excluded due to the hard-core interaction between spheres and counterions).

\begin{figure}[t]
\includegraphics[angle=0,scale=0.45]{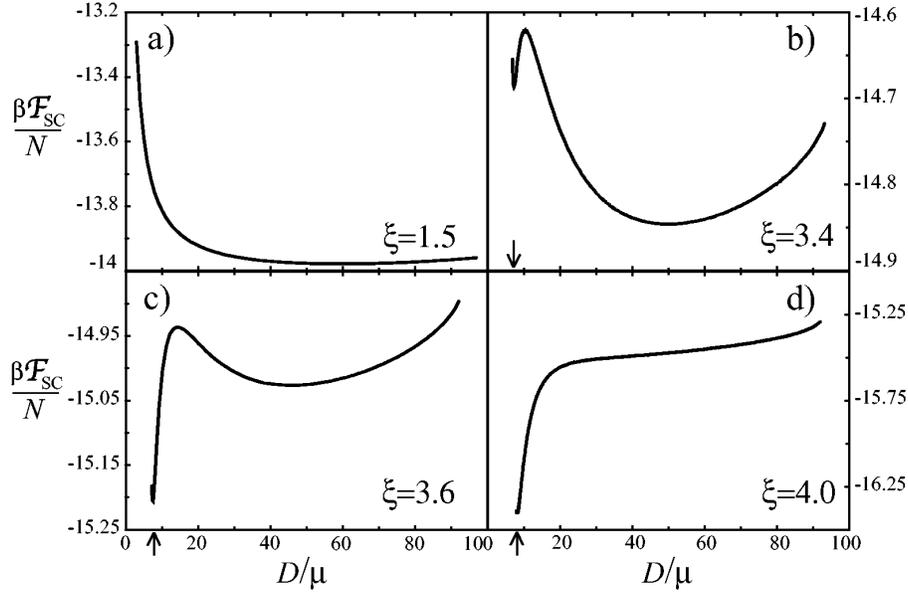}
\caption{The strong-coupling free energy of the two-sphere system, Eq. (\ref{eq:SCfree_sph}), plotted
as a function of 
the rescaled center-to-center distance, $D/\mu$, for Manning parameters a) $\xi=1.5$, b) $\xi=3.4$, c) $\xi=3.6$ 
and d) $\xi=4.0$ (the box size
is $L/\mu=100$). The location of the local minimum at small separations is marked by an arrow. A potential barrier
is found in the range of Manning parameter $\xi_c^{(1)}(L)<\xi<\xi_c^{(2)}(L)$, where 
we have  $\xi_c^{(1)}\approx 3.3$ and $\xi_c^{(2)}\approx 3.8$ for $L/\mu=100$.
The values of the two free energy minima become equal at $\xi\approx 3.5$.}
\label{fig:spheres_free}
\end{figure}

The energetic and entropic contributions from counterions enter on the leading order
through the second term in Eq. (\ref{eq:SCfree_sph}).  
It is important to note that the main qualitative features regarding the binding-unbinding 
behavior of counterions is reproduced by this term. 
In particular, for very large confining box $L\rightarrow \infty$, the single-particle partition function
${\mathcal Z}_1(D,\xi,L)= \int_V{\textrm{d}}^3 r \exp(-\beta u)$ diverges with
the box volume as ${\mathcal Z}_1\sim V=L^3$ for any given Manning parameter
(since the integrand is always positive 
and bigger than one). Thus the distribution of counterions around the spheres, 
$\sim\exp(-\beta u)/{\mathcal Z}_1$, as well as the component of the force contributed by counterions, 
$\sim \partial \ln {\mathcal Z}_1/\partial D\sim  L^{-2}$, vanish in the limit $L\rightarrow \infty$. 
This leads to a pure repulsion between unconfined spheres as expected. 
Note that this repulsion regime does not represent the strong-coupling situation
associated with large electrostatic correlations.
The fact that
both the de-condensation process and the repulsive 
regime for unconfined spheres are consistently captured by the asymptotic strong-coupling 
free energy indicates that the $1/\Xi$-expansion scheme used to  obtain the asymptotic contribution
for $\Xi\rightarrow\infty$ \cite{Netz01} is not only based on energetic considerations, 
and can also account for entropic effects on the leading order \cite{Netz01,Naji_epje04}.

Now let us consider the spheres in a finite confinement volume and investigate 
the strong-coupling prediction for the interaction free energy (\ref{eq:SCfree_sph}).  
For small Manning parameters, the free energy exhibits only a long-range repulsion 
(Figure \ref{fig:spheres_free}a), but as Manning parameter exceeds a threshold 
value of $\xi_c^{(1)}$, a local minimum is
developed at small separations indicating a short-range attraction and a meta-stable
bound state. As seen (Figures \ref{fig:spheres_free}b and c), 
this attraction regime is separated from the large-distance repulsion regime by 
a pronounced potential barrier. For increasing Manning parameter, the attractive local minimum 
becomes deeper than the large-distance minimum, 
and the potential barrier disappears beyond a second threshold of 
$\xi_c^{(2)}$--see Figure \ref{fig:spheres_free}d. 
The preceding features indicate a discontinuous unbinding transition between a closely-packed
bound-state and a repulsion-dominated state of two like-charged
spheres by varying the Manning parameter. 
The generic form of the free energy also agrees qualitatively 
with numerical findings \cite{Gron98} as discussed in Section \ref{subsec:spheres_sim}.

\begin{figure}[t]
\includegraphics[angle=0,scale=0.325]{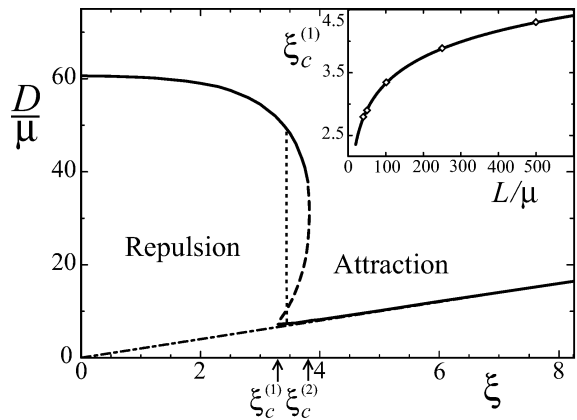}
\caption{Regimes of attraction and repulsion in the system of two like-charged 
spheres and counterions as obtained from the asymptotic strong-coupling 
theory. The solid curves show the rescaled equilibrium center-to-center distance between spheres
as a function of the 
single-sphere Manning parameter, $\xi$ (Eq. (\ref{eq:xi_sph})), for a
confining box of rescaled size $L/\mu=100$. The dashed curve corresponds 
to the maximum of the strong-coupling free energy and the dotted vertical line
shows the Manning parameter for which the values of the two minima of the 
free energy are equal (see Figure \ref{fig:spheres_free}). The locations of threshold Manning parameters
$\xi_c^{(1)}$ and $\xi_c^{(2)}$ are shown by arrows. The dot-dashed line
shows the contact separation $D=2R$.
Inset: The attraction threshold for two like-charged spheres, $\xi_c^{(1)}$,  increases weakly with the
rescaled box size, $L/\mu$. It exhibits a logarithmic dependence according to Eq. (\ref{eq:xic1}), which is 
shown by the solid curve.}
\label{fig:spheres_Dmin}
\end{figure}

A digram representing different regimes of attraction and repulsion is shown in 
Figure \ref{fig:spheres_Dmin},
where we have plotted the locations of the minima (solid curves) of the strong-coupling
free energy  (\ref{eq:SCfree_sph}) and also the location of its maximum 
(dashed curves) as a function of Manning parameter, $\xi$. (The locations of the two
threshold Manning parameters, $\xi_c^{(1)}$ and $\xi_c^{(2)}$, are shown by arrows.)
One can show  that for $\xi\ll 1$, the
location of the repulsion-dominated minimum at large separations
(Figure  \ref{fig:spheres_free}a), $D_\ast$, scales linearly 
with the box size as 
\begin{equation}
  D_\ast\approx \sqrt[3]{\frac{3}{4\pi}}  L,
\label{eq:D0sph}
\end{equation} 
when the box size tends to infinity $L\rightarrow \infty$ \cite{Naji_epje04}. 
On the other hand, for large Manning parameter $\xi\gg 1$, the location of the 
attraction-dominated minimum at small separations (Figure  \ref{fig:spheres_free}d), $D_\ast$, 
saturates to a value independent from the box size; in this case, the  
equilibrium surface-to-surface distance
of spheres, $\Delta_\ast\equiv D_\ast-2R$,  follows from Eq. (\ref{eq:SCfree_sph}) 
approximately as 
\begin{equation}
  \Delta_\ast\equiv D_\ast-2R\approx \sigma_{\textrm{ci}}+\frac{4}{7}\mu+{\mathcal O}(\mu^2),
\label{eq:Delta_sph}
\end{equation}
where $\sigma_{\textrm{ci}}$ is the counterion diameter. This minimum corresponds 
to a highly-condensed state of counterions in the intervening region between spheres \cite{Naji_epje04}. 
The attractive force induced between spheres in this regime ($\xi\gg 1$) may also be calculated
from the strong-coupling free energy (\ref{eq:SCfree_sph}); 
at small separations ($D\sim 2R$), the force takes the following analytical form 
\begin{equation}
   F(D)\approx -7\frac{Z^2e^2}{4\pi \varepsilon \varepsilon_0 D^2}.
\label{eq:f_sph}
\end{equation}
This limiting attractive force (for $\Xi\gg 1$ and $\xi\gg 1$) is independent from the temperature and 
results only from energetic contributions \cite{Naji_epje04}. 
The expression (\ref{eq:f_sph}) qualitatively agrees with the 
results obtained by Shklovskii using the Wigner-crystal model  \cite{Shklovs99}. 

The predictions of the strong-coupling theory for 
the bound-state separation of two attracting spheres, $\Delta_\ast$,
has been compared with numerical simulations in Table \ref{tab:sim_parameters_sph}
(note that here $\Delta_\ast$ has been calculated by numerical evaluation of the free energy 
(\ref{eq:SCfree_sph}) for the corresponding simulation parameters. The analytical
expression (\ref{eq:Delta_sph}) may also be used, but it gives an approximate 
value up to the first order in $\mu$).   
As seen there is a reasonable semi-quantitative agreement between the theoretical predictions 
and the simulation results. 
As mentioned in Section \ref{subsec:spheres_sim}, the equilibrium surface-to-surface distance
in these simulations appears to be about the counterion diameter (see the Discussion
in Ref. \cite{Linse00}). This also follows from the strong-coupling prediction, Eq. (\ref{eq:Delta_sph}), since
for highly-charged spheres, the Gouy-Chapman length
is in fact small compared with the counterion diameter. 
One should also note that the lateral separation of counterions at spheres in the simulations
is typically larger than the counterion diameter  indicating that the excluded-volume 
interaction between counterions is not a dominant effect. The volume interactions
between counterions enter only in the higher-order corrections
to the asymptotic theory and can lead to additional attractive
 components between macroions \cite{AllahyarovPRL,Deserno03}.
 
\subsection{Attraction threshold}
\label{subsec:sph_threshold}

The strong-coupling theory also allows to obtain an analytical 
estimate for the regime of Manning parameters where attraction is expected between spheres.
This regime may be specified by
\begin{equation}
  \xi>\xi_c^{(1)},
\label{eq:xi_crit}
 \end{equation}
where the threshold Manning parameter $\xi_c^{(1)}$ actually depends on the confinement size, 
$\xi_c^{(1)}=\xi_c^{(1)}(L)$.  As shown in the inset of
 Figure \ref{fig:spheres_Dmin}, $\xi_c^{(1)}(L)$ increases almost
logarithmically with the box size as
\begin{equation}
  \xi_c^{(1)}(L)\approx a+b\ln \left(\frac{L}{\mu}\right),
\label{eq:xic1}
\end{equation} 
where $a\approx 0.55$ and $b\approx 0.6$ are obtained by 
fitting to numerically-determined SC predictions (symbols). 
The estimated values of $\xi_c^{(1)}$ are shown in Table  \ref{tab:sim_parameters_sph}
for the given simulation parameters, which show that these simulations indeed exhibit
the attraction regime  (\ref{eq:xi_crit}). 
The weak dependence of the attraction threshold, $\xi_c^{(1)}$,
on the confinement size can also explain the stability of compact clusters
of spheres in quite large confinements ($L/\mu\gg 1$) \cite{Gron98,Wu,AllahyarovPRL,LinsePRL,Linse00,Hribar,Messina00}, 
since the Manning parameter only needs to exceed a moderate value ($\sim \ln L/\mu$)
for like-charged spheres to fall into the attraction-dominated regime. 

\begin{figure}[t]
\includegraphics[angle=0,scale=0.6]{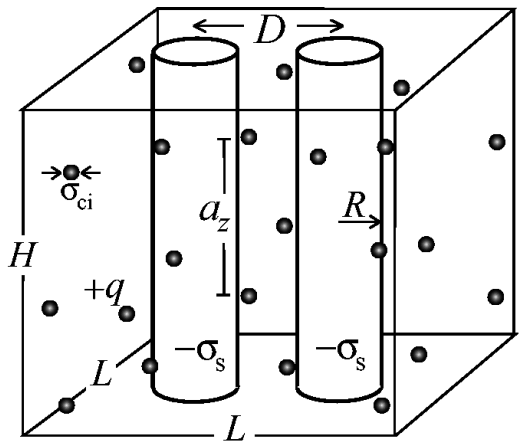}
\caption{Two identical and parallel charged cylinders of radius $R$ are considered at axial 
separation of $D$ and in a square box of lateral edge size $L$; cylinders 
have a length of $H$, which is assumed to be infinitely large. The bare 
charge of cylinders with uniform surface charge density $-\sigma_{\textrm{s}}$ (in units
of the elementary charge, $e$, and assumed to
be negative) is 
compensated by the total charge of counterions of charge valency $q$ (and diameter
$\sigma_{\textrm{ci}}$). }
\label{fig:rods_model}
\end{figure}

\section{Attraction between like-charged cylinders}
\label{sec:rods}

\subsection{Effective cylinder-cylinder interaction: Asymptotic strong-coupling theory}
\label{subsec:rods_SC}

In this Section, we shall consider the interaction between two like-charged cylinders 
in the limit of large coupling parameter $\Xi\rightarrow\infty$. 
For simplicity, we assume that the cylinders are infinitely long (with length $H$), and that they are
 confined in a square box of edge size $L$ (see Figure \ref{fig:rods_model}).
 The electroneutrality condition holds  inside the confining box, thus we have
$q N=2 \tau H$, where $N$ is the number of counterions and $\tau=2\pi \sigma_{\textrm{s}} R$ is 
the linear charge density of each cylinder. We assume that counterions and cylinders
have also a hard-core excluded-volume interaction. 

The strong-coupling free energy of this system (up to an irrelevant additive term) 
follows as \cite{Naji_epje04} 
\begin{equation}
  \frac{\beta {\mathcal F}_{\textrm{SC}}}{N}=-\xi \ln \tilde D-
                 \ln \int_V\,{\textrm{d}}x\,{\textrm{d}}y\, e^{-\beta u(x,y)}, 
\label{eq:SCfree_rods}
\end{equation}
where $\tilde D=D/\mu$ is the axial separation of the cylinders in units of the Gouy-Chapman
length. The first term in Eq. (\ref{eq:SCfree_rods})
is the bare repulsion of cylinders and in the second term,
$ \beta u$ is the single-particle interaction energy of counterions with the cylinders, 
\begin{equation} 
          \beta u= 2\xi
                            (\ln {\tilde r}_1+\ln {\tilde r}_2),
\end{equation}    
where $\tilde r_1=r_1/\mu$ and $\tilde r_2=r_2/\mu$ are the radial distances 
 from the axes of the two cylinders (in the $xy$-plane perpendicular
to the cylinders axes)--we choose the frame of reference in the middle of the box
such that $r_{1,2}=[(x\pm D/2)^2+y^2]^{1/2}$. The integral in Eq. (\ref{eq:SCfree_rods}) 
runs over the volume accessible for counterions inside the box excluding the two cylinders
(and a shell of thickness $\sigma_{\textrm{ci}}/2$ around each cylinder corresponding
to the closest approach distance of counterions). 

As in the case of two spheres (Section \ref{subsec:spheres_SC}), 
the effective interaction between 
like-charged cylinders is influenced by the binding-unbinding
behavior of counterions incorporated on the leading order in the single-particle partition function
${\mathcal Z}_1(D,\xi,L)= \int_V{\textrm{d}}^2 r \exp(-\beta u)$ in the second term of the
free energy (\ref{eq:SCfree_rods}).
For further analysis of this behavior in the two-cylinder system, let us first consider 
the limit of very large box size $L\rightarrow \infty$. 
In this limit, ${\mathcal Z}_1$ scales with the box size as ${\mathcal Z}_1\sim L^{2-4\xi}$, 
which may be seen simply by rescaling the spatial coordinates with $L$ 
as $x\rightarrow x/L$, {\em etc}. 
Thus for $\xi<1/2$, ${\mathcal Z}_1$ diverges and consequently, the distribution function
of counterions, $\sim \exp(-\beta u)/{\mathcal Z}_1$, vanishes indicating  
de-condensation of counterions from the two cylinders. 
The counterion-mediated force between 
cylinders, $\sim \partial \ln {\mathcal Z}_1/\partial D$, tends to zero as well; thus 
the cylinders only repel each other in the limit $L\rightarrow \infty$. 
In contrast, for Manning parameter $\xi>1/2$, counterions bind to the 
two unconfined cylinders with a finite density profile, and thus can 
produce attraction between them. 
The condensation threshold obtained above for two cylinders, $\xi_\ast=1/2$,
agrees with the classical result due to Manning  \cite{Manning97} quantitatively.

An interesting question is to obtain the regime of Manning parameters, where
the attraction emerges between cylinders. Intuitively, we expect that the effective 
attraction sets in somewhat above the counterion-condensation threshold $\xi_\ast=1/2$.\
Because right at the condensation threshold, there is an unbalanced bare repulsion between 
the cylinders, and a finite fraction of condensed counterions is needed to compensate this repulsion.  
It follows from a more detailed analysis of the strong-coupling 
free energy (\ref{eq:SCfree_rods})
that the attraction actually emerges when  \cite{Naji_epje04}
\begin{equation}
  \xi>\xi_c=2/3,
\end{equation}
which gives a universal threshold for attraction between unconfined cylinders.

The attraction threshold between two like-charged cylinders has also been investigated using other 
methods. 
Analysis of Ray and Manning \cite{Manning97} based on the classical counterion-condensation 
theory \cite{Manning69} predicts attraction between two like-charged cylinders for $\xi>1/2$, which coincides
with the onset of counterion condensation. It should be noted, however, that 
the attraction mechanism involved in their theory is not based on electrostatic 
correlations, but features a mean-field covalence-like binding process. 
Arenzon {\em et al.}'s study \cite{Arenzon99,Levin02} based on a structural-correlations theory
(which also accounts for counterion condensation) predicts attraction for  $\xi>2$. Numerical simulations 
\cite{Gron97,Deserno03,Naji_epl04,Lee04}, on the other hand, give
attraction for the range of Manning parameters $\xi>0.8$, but have not 
yet specified the attraction threshold precisely (Section \ref{subsec:rods_sim}).

\begin{figure}[t]
\includegraphics[angle=0,scale=0.325]{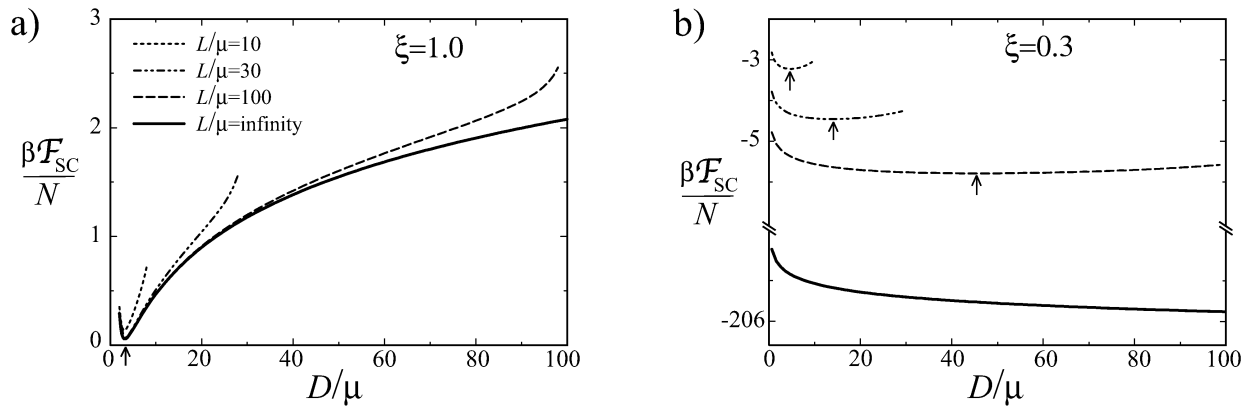}
\caption{The strong-coupling free energy of the two-cylinder system as a function of the rescaled axial
distance, $D/\mu$,  for Manning parameters a) $\xi=1.0$ and b) $\xi=0.3$ and for  several box sizes as 
indicated on the graph.
Arrows show the approximate location of the minimum of the free energy, which for large Manning parameters is
nearly independent of the box size and reflects a closely-packed bound state. For small Manning parameter
$\xi<\xi_c=2/3$, the location of the minimum tends to infinity with the box size reflecting a continuous unbinding 
transition for like-charged cylinders.}
\label{fig:rods_free}
\end{figure}

For two cylinders in a finite confinement volume, one can determine the
attraction and repulsion regimes by evaluating the strong-coupling
free energy Eq. (\ref{eq:SCfree_rods}). The typical form of the 
free energy  (\ref{eq:SCfree_rods})  is shown in Figure \ref{fig:rods_free} for
both large ($\xi>\xi_c=2/3$) and small ($\xi<\xi_c=2/3$) Manning parameters. 
For large Manning parameter (Figure \ref{fig:rods_free}a), the free energy exhibits a long-range attraction and a local
minimum at small separations, which is quite insensitive to
the confinement size as counterions are localized mostly in the proximity of the cylinders
in this regime. The analytical form of the 
attraction force (per unit length of the cylinders $H$) 
may be estimated for large Manning parameters $\xi\gg 1$ \cite{Naji_epje04};
in the zero-temperature limit ($\xi\rightarrow \infty$), we have:
\begin{equation}
      \frac{F(D)}{H}\approx 
           -\frac{e^2\tau^2}{2\pi \varepsilon \varepsilon_0}\times \left\{
             \begin{array}{ll}
               1/D
              & {\,\,\,\,\,\,\,\,\,D\gg 2R,}\\
            \\
              3/D 
              & {\,\,\,\,\,\,\,\,\,D\approx 2R,}
      \end{array}
        \right. 
\label{eq:SCforce_rods}
\end{equation}
for large and small axial separations respectively. 
This limiting attractive force (which is obtained for $\Xi\gg 1$ and $\xi\gg 1$)
is independent from the temperature and originates from a purely electrostatic origin. 

For small Manning parameter $\xi<\xi_c=2/3$  (Figure \ref{fig:rods_free}b), 
the interaction free energy of cylinders is dominated by their bare repulsion
as counterionic clouds become increasingly diluted around the cylinders. 
The location of the minimum of the free energy for $\xi\ll 1$, $D_\ast$, tends to infinity 
with the box size as $D_\ast\approx L/\sqrt{\pi}$. 

\begin{figure}[t]
\includegraphics[angle=0,scale=0.325]{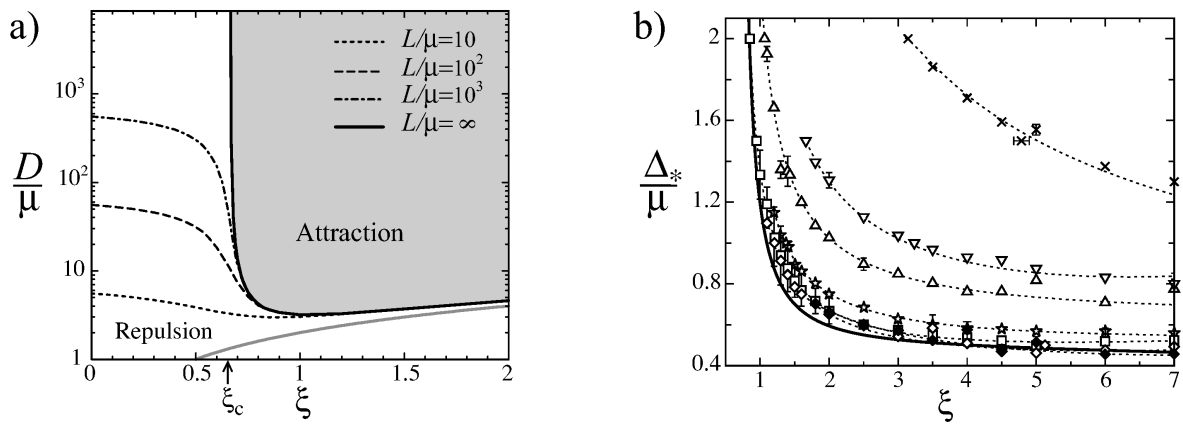}
\caption{a) Regimes of attraction and repulsion in the system of two like-charged 
cylinders and counterions as obtained from the asymptotic strong-coupling 
theory. The curves show the rescaled equilibrium axial separation between
cylinders, $D_\ast/\mu$, as a function the 
single-cylinder Manning parameter, $\xi$ (Eq. (\ref{eq:xi_rods})), for 
confining boxes of different size as indicated on the graph. The curves
are obtained numerically by minimization of the free energy (\ref{eq:SCfree_rods}).
In an unconfined box, the cylinders continuously unbind at the threshold Manning parameter
$\xi_c=2/3$ (shown by an arrow). The thick gray curve corresponds to 
$D=2R$, where the two cylinders are at contact.\\
b) The rescaled equilibrium surface-to-surface separation of the cylinders, 
$\Delta_\ast/\mu=(D_\ast-2R)/\mu$, as a function of Manning parameter, $\xi$. Symbols are
simulation data for the Rouzina-Bloomfield parameter (from top):
$\gamma_{\textrm{RB}}=3$ (crosses), 10 (triangle-downs), 15 (triangle-ups),
30 (stars), 40 (squares), 50 (open diamonds) and 60 (filled diamonds) \cite{Naji_epl04}.
The solid curve is the strong-coupling prediction obtained from Eq.  (\ref{eq:SCfree_rods}). 
Dashed curves are guides to the eye.}
\label{fig:rods_Dmin}
\end{figure}

In brief, one may specify the attraction and repulsion regimes of two like-charged cylinders
by considering the location of the minimum of the 
free energy (\ref{eq:SCfree_rods}) as shown in Figure \ref{fig:rods_Dmin}a 
for several different box sizes
(the region below each curve shows the repulsion regime and above 
that is the attraction regime). 
For $\xi\gg 1$, the equilibrium surface-to-surface
separation of the cylinders, $\Delta_\ast$, may be obtained approximately as
\begin{equation}
  \Delta_\ast\equiv D_\ast-2R\approx \sigma_{\textrm{ci}}+\frac{2}{3}\mu+{\mathcal O}(\mu^2),
\label{eq:Delta_rods}
\end{equation}
where $\sigma_{\textrm{ci}}$ is the counterion diameter. For typical strongly-coupled systems (with
small Gouy-Chapman length), the equilibrium surface separation is predicted from 
Eq. (\ref{eq:Delta_rods}) to be of the order of the counterion diameter. 

When Manning parameter decreases down to the attraction threshold 
$\xi_c=2/3$, the two cylinders unbind from each other for $L\rightarrow\infty$ (solid curve in Figure \ref{fig:rods_Dmin}a). 
But in contrast to like-charged spheres, the unbinding transition of cylinders occurs 
in a continuous fashion exhibiting a universal scaling exponent for the
diverging axial distance of cylinders  \cite{Naji_epje04}
\begin{equation}
    D_\ast\sim (\xi-\xi_c)^{-\alpha},
\label{eq:scaling_rods}
\end{equation} 
where the exponent is found as 
\begin{equation}
   \alpha = 3/2.
\end{equation}

\subsection{Comparison with numerical simulations}
\label{subsec:rods_sim}

Several numerical simulations have been reported on charged cylinders 
\cite{Gron97,Deserno03,Naji_epl04,Lee04}
and also on more detailed models, which incorporate the charge pattern of DNA \cite{Lyub95,Allahyarov00}.
Attraction is found in a wide range of Manning parameters (including $\xi\approx 1.0$)
and for moderate to large coupling parameters. 

Gr{\o}nbech-Jensen {\em et al.} \cite{Gron97} 
showed that the effective interaction between two parallel like-charged
cylinders  in the presence of divalent counterions exhibits repulsion at very 
small distances close to contact and 
attraction at intermediate distances by means of which a closely-packed bound state is maintained at
small surface separations. At large distances the effective interaction turns to repulsion. 
Similar results have been reported in simulations by Lee {\em et al.} \cite{Lee04}.
In these simulations, charges on cylinders and counterions interact with Coulombic
interactions as well as short-range excluded-volume interactions.  Recent simulations by 
Deserno {\em et. al} \cite{Deserno03} indicate an interplay between 
electrostatic and excluded-volume interactions, which may 
drastically influence the effective {\em electrostatic}
interaction between like-charged cylinders. Specially for large linear charge densities
(when charge separation on the cylinder is smaller than the counterion size), the attractive electrostatic
force between cylinders weakens and eventually turns to repulsion for sufficiently large coupling  
strength \cite{Deserno03}.  In this case, the {\em total} effective force between 
the cylinders is still attractive due to an
attractive component from excluded-volume interactions. 
This behavior is not yet completely understood and
appears to be different from a simple depletion mechanism \cite{Deserno03}. 
A similar situation may be present
in the system studied in Ref. \cite{Gron97}, however, the detailed analysis of the force components is not
reported.

Intuitively, we expect that excluded-volume effects become important when typical distance
between counterions on cylinders surface, $a_z=q/\tau$ (Section \ref{subsubsec:RB_rods}),
is smaller than the counterion diameter, $\sigma_{\textrm{ci}}$, {\em i.e.} for $a_z<\sigma_{\textrm{ci}}$. 
In this case, excess accumulation of counterions in the intervening 
region between cylinders, which is favored energetically and leads to the strong-coupling attraction
\cite{Naji_epje04}, is prohibited \cite{Deserno03}.  
In order to investigate the electrostatic features of like-charge attraction in the two-cylinder
system and to avoid complications arising from volume interactions, we consider recent Molecular
Dynamics simulations performed by Arnold and Holm, which exclude volume interactions between 
counterions  though counterions still retain a soft-core excluded-volume repulsion with the cylinders \cite{Naji_epl04}. 
This is important for the forthcoming comparison made with the strong-coupling predictions, 
since the leading-order results at large couplings only involve counterion-cylinder interactions
(Section \ref{sec:rods}).   
The simulation model has a geometry similar to what we have shown in Figure \ref{fig:rods_model},
where periodic boundary conditions in $z$ direction is used. (For convenience the simulated model 
employs a cylindrical outer boundary of diameter $8D$. For final comparison, the theoretical curves are also
calculated using a similar constraint, though for simplicity, a square box of edge size
$L=8D$ is used, which as explicitly checked does
not influence the following results in the considered range of Manning parameters \cite{Naji_epl04}.)
The soft-core repulsion in the simulations is also chosen 
strong enough such that it effectively prevents counterions from penetrating the cylinders, which makes the 
simulation model comparable to the theoretical model  with a hard-core volume interaction.
In the simulations, cylinders are kept at fixed surface-to-surface distance, $\Delta$,
with fixed linear charge density, $\tau$,  and counterion valency, $q$, but the Bjerrum length and
the cylinders radius, $R$, are varied. Hence, the Gouy-Chapman length, $\mu$ (Eq. (\ref{eq:mu_curve})),
varies accordingly allowing to span the phase space of the system and specify the attraction and repulsion regimes.

In Figure \ref{fig:rods_Dmin}b, we show the simulation results for the equilibrium 
surface-to-surface distance of cylinders in rescaled
units, $\Delta_\ast/\mu$, as a function of Manning parameter $\xi=q \ell_{\textrm{B}} \tau$ 
(Eq. (\ref{eq:xi_rods})). Different symbols correspond to different values of the
{\em Rouzina-Bloomfield parameter}  defined as 
\begin{equation}
   \gamma_{\textrm{RB}}=\frac{q}{\tau \Delta},
\end{equation}
which is fixed for each simulation data set. This dimensionless ratio gives a measure of the deviations from
the strong-coupling regime, since it represents the ratio between the estimated correlation hole
size, $a_z=q/\tau$ \cite{Deserno03}, and the surface-to-surface distance between cylinders 
(see Section  \ref{subsubsec:RB_rods})  \cite{Note_gammaRB}.
Thus according to Rouzina-Bloomfield criterion, large $\gamma_{\textrm{RB}}=a_z/\Delta$ corresponds to
a strongly-coupled system with highly correlated structures on opposite surfaces of cylinders that may
give rise to attraction between them.  
This behavior is clearly supported by the present simulations as seen in Figure \ref{fig:rods_Dmin}b. 
For increasing $\gamma_{\textrm{RB}}$ (from about 3 up to 60),  the  equilibrium separation between 
cylinders decreases and tends to the strong-coupling prediction indicating a closely-packed 
bound state at large Manning parameters $\xi>1$, where the equilibrium surface-to-surface 
distance is of the order of the Gouy-Chapman length.  

The quantitative agreement
with the strong-coupling prediction is obtained for the whole range of Manning parameters 
studied in the simulations. Due to convergence limitations, the simulations so far have been 
limited to the range of Manning parameters $\xi>0.8$. It becomes more difficult to obtain good 
data for smaller Manning parameters as the distance between cylinders rapidly increases.
Nonetheless, the excellent convergence of the present data to the strong-coupling curve
suggests an attraction threshold of about $\xi_c=2/3$ obtained in Section \ref{subsec:rods_SC}.

%
%

\section{Concluding remarks and Discussion}

We have discussed the effective electrostatic 
interaction between like-charged macroions in the regime of large coupling 
parameter, $\Xi$, which is achieved for large counterion valency, large charge densities
on macroions, at low dielectric constants or low temperatures. 
In this regime, interactions between macroions are dominated by 
strong counterionic correlations: counterions form highly correlated layers at
macroions, including Wigner crystals for sufficiently large coupling, that 
lead to attractive forces between like-charged macroions at sufficiently 
small surface separations. The relevant length scale is set by the typical distance between  
counterions at macroion surfaces (the correlation hole size) below which like-charged
surfaces couple to each other electrostatically. This energetic coupling is essentially mediated
by single counterions sandwiched between opposite surfaces of macroions. 
At larger surface separations, however, correlation effects  and the attraction strength
are reduced and multi-counterion interactions play a significant role.
At very large separation (as long as the coupling parameter is finite), the interaction 
between macroions is dominated by mean-field features and eventually turns to
repulsion. 

The strong-coupling attraction appears with the same mechanism for planar and 
curved surfaces. For curved surfaces, however, attraction is found only when a sufficiently
large number of counterions are condensed near the surface; the condensation process is
regulated by the ratio between the radius of curvature, $R$, and the Gouy-Chapman length, $\mu$,
that is the Manning parameter $\xi=R/\mu$.  Therefore, the attraction regime for charged 
cylinders and spheres is characterized by both a large coupling parameter $\Xi$, which is specified
via the Rouzina-Bloomfield criterion, and also by a large Manning parameter, which has to be
determined by considering the condensation process of counterions in the given system of macroions
(Sections \ref{sec:curve}-\ref{sec:rods}). 

We reviewed recent analytical results obtained in the limit of $\Xi\rightarrow \infty$,
which gives rise to the asymptotic strong-coupling theory. This theory incorporates single-particle
contributions on a systematic level 
and may be extended to finite coupling regimes by considering higher-order corrections
via a $1/\Xi$-expansion series (virial expansion) \cite{Netz01}.  
The predictions of the strong-coupling theory were considered for the effective interaction between 
two charged walls and also for two spherical and cylindrical macroions.
This asymptotic theory yields a long-ranged attractive force  
between like-charged objects that is of constant strength for two walls, but varies
with distance, $D$, as  $\sim 1/D$ for two cylinders and as $\sim 1/D^2$ for two spheres.
Note that the leading-order free energy 
contains in addition to energetic contributions also (repulsive) entropic 
 contributions from
counterions (see Eqs. (\ref{eq:P_SC}), (\ref{eq:SCfree_sph}) and (\ref{eq:SCfree_rods})), 
which vanish  at  zero temperature.
The zero temperature limit is in fact contained within the strong-coupling theory and may 
be obtained by taking the limit $\mu\rightarrow 0$ (or  $\xi\rightarrow \infty$ for curved surfaces), 
where the free energy reduces to a purely energetic contribution giving a long-ranged attractive force 
as specified above (see Eqs. (\ref{eq:P_SC}), (\ref{eq:f_sph}), (\ref{eq:SCforce_rods}) and Ref. \cite{Naji_epje04}). 
As we know from analytical and numerical studies of planar walls \cite{AndreEPJE,Netz01}, 
the range of strong-coupling attraction is reduced substantially at finite couplings due to
higher-order corrections. (The higher-order corrections have not yet been calculated for
the system of two cylinders and two spheres.)  By comparing the strong coupling predictions
with recent simulations, it was shown that the attraction regime and the closely-packed
bound state of macroions can be described by this limiting theory 
on a quantitative level \cite{AndreEPL,AndrePRL,AndreEPJE,Naji_epje04,Naji_epl04}. 
Note that the agreement with simulations is limited to the regime of surface separations determined by
the Rouzina-Bloomfield criterion. This regime particularly includes typical
coupling parameters of $\Xi\sim 10^2$ and Manning parameters of
$\xi\sim 10$ (for spheres) and $\xi>1$ (for cylinders), which are accessible 
in usual experimental systems (see Tables \ref{tab:real_parameters} 
and \ref{tab:sim_parameters_sph}).  

The asymptotic strong-coupling theory was also considered  
to estimate the threshold Manning parameter above which the attraction is expected for two 
like-charged spheres and cylinders. The attraction threshold is captured within the asymptotic theory
because it accounts for the entropy-driven de-condensation 
process of counterions on the leading order. One should note, however, that the de-condensation regime
does not involve electrostatic correlations and thus in general, the strong-coupling theory may
be only qualitatively valid when counterions de-condense at {\em low} Manning parameters. 
Yet strong-coupling predictions were shown to remain in a reasonable  agreement 
with simulations on two cylinders for decreasing Manning parameter close to de-condensation 
threshold \cite{Naji_epl04}. (Moreover,  the predicted de-condensation threshold itself 
agrees quantitatively with the standard Manning results for one and two cylinders \cite{Manning69,Manning97}.)
Further numerical and analytical studies are useful to specify the
validity of the strong-coupling predictions at low Manning parameters, including the 
predicted unbinding behavior of cylinders and spheres. 

We did not discuss possible thermodynamic phase transitions triggered by attractive forces in the system
of charged plates, spheres and cylinders. (Note that the binding-unbinding behaviors discussed
in Sections  \ref{sec:spheres} and \ref{sec:rods} represent thermodynamic phase transitions
only in the case of infinitely long cylinders; for spheres (or short cylinders) there will be 
additional entropic contributions to the free energy from the sphere-sphere distance coordinate, 
which are not considered within the present model.)
A first-order unbinding transition was predicted to occur in the system of two like-charged
walls at the coupling parameter $\Xi\approx 17$ that has also been compared with experimental 
observations \cite{AndreEPJE}. Also there has been indication of an attraction-induced
 phase separation in the system of like-charged spheres
from recent numerical simulations \cite{Gron97,LinsePRL,Linse00}. 
The systematic study of such phase transitions still remains a challenging subject (see Ref. \cite{DBL}
and references therein).  

Another interesting problem is to examine the influence of additional salt on the 
interactions in the strong-coupling limit. The present results are expected to remain valid 
at sufficiently small salt concentrations (large screening length). Qualitatively, one can associate the
size of the confinement box considered in the present models with the screening length. Thus, addition of  salt 
is expected to matter particularly close to the unbinding threshold of macroions.
Other relevant subjects include the effect of finite polymer 
stiffness, the discrete charge pattern of macroions \cite{Lyub95,Allahyarov00,Gonz01,Korny,Arenzon99,Shklovs99,Diehl01,Andre02}, 
and bundling of many charged polymers  \cite{Bloom,Tang,Stevens99,Ha,Shklovs99,Shklo00}
in the strong coupling regime, which constitute interesting applications for the future. 



\begin{acknowledgments}
We acknowledge financial supports from the DFG Schwerpunkt Polyelektrolytes with defined architecture
and the DFG German-French Network.
\end{acknowledgments}




\begin{thebibliography}{21}


\bibitem{VO} E.J. Verwey, J.T.G. Overbeek, {\em Theory of the Stability of 
Lyophobic Colloids} (Elsevier, Amsterdam, 1948).

\bibitem{Israelachvili}
J. Israelachvili, {\em Intermolecular and Surface Forces} (Academic Press, London, 1991). 

\bibitem{Hunter} 
R.J. Hunter, {\em Foundations of Colloidal Science} (Clarendon, Oxford, 1987). 

\bibitem{Napper}
D.H. Napper, {\em Polymeric Stabilization of Colloidal Dispersions} (Academic Press, London, 1983). 

\bibitem{Polyelec}
H. Dautzenberg, W. Jaeger, B.P.J. K\"otz, C. Seidel, D. Stscherbina, {\em Polyelectrolytes: Formation, 
characterization and application} (Hanser Publishers, New York, 1994). 

\bibitem{Yager}
T.D. Yager, C.T. McMurray, K.E. van Holde, Biochem. {\bf 23}, 2271 (1989). 

\bibitem{Khan}
A. Khan, B. J{\"o}nsson,  H. Wennerstr{\"o}m,  J. Phys. Chem. {\bf 89},  5180  (1985).

\bibitem{Wennerstroem91}
H. Wennerstr{\"o}m, A. Khan,  B. Lindman, Adv.\ Colloid Interface Sci. {\bf 34},  433  (1991).

\bibitem{Quirk}
R. Kjellander, S. Mar{\v{c}}elja,  J.P. Quirk, J. Colloid Interface Sci. {\bf 126}, 194 (1988).


\bibitem{Bloom} V.A.  Bloomfield, Biopolymers {\bf 31}, 1471 (1991); 
Curr. Opin. Struct. Biol. {\bf 6}, 334 (1996). 

\bibitem{Kekicheff93}
P. K{\'e}kicheff, S. Mar{\v{c}}elja, T.~J. Senden,  V.~E. Shubin, J.\ Chem.\
  Phys. {\bf 99},  6098  (1993).


\bibitem{Delsanti} M. Delsanti, J.P. Dalbiez, O. Spalla, L. Belloni, 
M. Drifford, ACS Symp. Ser. {\bf 548}, 381 (1994).

\bibitem{Della} P. Gonzalez-Monzuelos, M. Olvera de la Cruz, J. Chem. 
Phys. {\bf 103}, 3145 (1995); M. Olvera de la Cruz, L. Belloni, M. Delsanti, 
J.P. Dalbiez, O. Spalla, M. Drifford, J. Chem. Phys. {\bf 103}, 5781 (1995). 

\bibitem{Dubois}
M. Dubois, T. Zemb, N. Fuller, R.P. Rand,  V.A. Parsegian, J. Chem. Phys. 
{\bf 108}, 7855 (1998).

\bibitem{Strey98} H.H. Strey, R. Podgornik, D.C. Rau, V.A. Parsegian, 
Curr. Opin. Struct. Biol. {\bf 8}, 309 (1998).

\bibitem{Tang} J.X. Tang, S. Wong, P. Tran, P.A. Janmey,
Ber. Bunsen-Ges. Phys. Chem. {\bf 100}, 1 (1996); J.X. Tang, T. Ito, T. Tao, 
P. Traub, P.A. Janmey, Biochemistry {\bf 36}, 12600 (1997). 

\bibitem{Tang03} G.C.L. Wong, A. Lin, J.X. Tang, Y. Li, P.A. Janmey, 
C.R. Safinya, Phys. Rev. Lett. {\bf 91}, 018103 (2003); J.C. Butler, 
T. Angelini, J.X. Tang, G.C.L. Wong, Phys. Rev. Lett. {\bf 91}, 
028301 (2003).

\bibitem{Angelini03}
T.E. Angelini, H. Liang, W. Wriggers, G.C.L. Wong, Proc. Natl. Acad. Sci. USA {\bf 100},
8634 (2003)

\bibitem{Guld84} L. Guldbrand, B. J\"onsson, H. Wennerstr\"om, P. Linse,
J. Chem. Phys. {\bf 80}, 2221 (1984). 

\bibitem{Svensson} B. Svensson, B. J\"onsson, Chem. Phys. Lett. {\bf 108}, 
580 (1984). 

\bibitem{Bratko86}
D. Bratko, B. J{\" o}nsson,  H. Wennerstr{\" o}m, Chem.\ Phys.\ Lett. 
{\bf  128},  449  (1986).

\bibitem{Guld86} L. Guldbrand, L.G. Nilsson,  L. Nordenski\"old, 
J. Chem. Phys. {\bf 85}, 6686 (1986); L.G. Nilsson, L. Guldbrand, 
L. Nordenski\"old, Mol. Phys. {\bf 72}, 177 (1991). 

\bibitem{Wood88} C.E. Woodward, B. J\"onsson, T. {\AA}kesson, 
J. Chem. Phys. {\bf 89}, 5145 (1988). 

\bibitem{Valleau} J.P. Valleau, R. Ivkov, G.M. Torrie, J. Chem. Phys. 
{\bf 95}, 520 (1991). 

\bibitem{Kjellander92} R. Kjellander, T. {\AA}kesson, 
B. J{\" o}nsson, S. Mar{\v{c}}elja, J. Chem. Phys. {\bf 97},  
1424 (1992).

\bibitem{Lyub95} A.P. Lyubartsev, L. Nordenski\"old, J. Phys. Chem. 
{\bf 99}, 10373 (1995).   

\bibitem{Gron97} N. Gr{\o}nbech-Jensen, R.J. Mashl, R.F. Bruinsma,
W.M. Gelbart, Phys. Rev. Lett. {\bf 78}, 2477 (1997). 

\bibitem{Gron98}  N. Gr{\o}nbech-Jensen, K.M. Beardmore, P. Pincus, 
Physica A {\bf 261}, 74 (1998). 

\bibitem{Wu} J. Wu, D. Bratko, J.M. Prausnitz, Proc. Natl. Acad. 
Sci. USA {\bf 95}, 15169 (1998); J. Wu, D. Bratko, H.W. Blanch, J.M. Prausnitz,
J. Chem. Phys. {\bf 111}, 7084 (1999).

\bibitem{AllahyarovPRL} E. Allahyarov, I. D'Amico, H. L\"owen, 
Phys. Rev. Lett. {\bf 81}, 1334 (1998). 

\bibitem{Stevens99} M.J. Stevens, Phys. Rev. Lett. {\bf 82}, 101 (1999). 

\bibitem{LinsePRL} P. Linse, V. Lobaskin, Phys. Rev. Lett. {\bf 83}, 
4208 (1999). 

\bibitem{Linse00} P. Linse, V. Lobaskin, J. Chem. Phys. {\bf 112}, 3917 
(2000); P. Linse, J. Chem. Phys. {\bf 113}, 4359 (2000). 

\bibitem{Hribar} B. Hribar, V. Vlachy, Biophys. J. {\bf 78}, 694 
(2000). 

\bibitem{Messina00}  R. Messina, C. Holm, K. Kremer, Phys. Rev. Lett. 
{\bf 85}, 872 (2000); Europhys. Lett. {\bf 51}, 461 (2000).

\bibitem{Allahyarov00} E. Allahyarov, H. L\"owen, Phys. Rev. E
{\bf 62}, 5542 (2000).  

\bibitem{AndreEPL}
A.G. Moreira, R.R. Netz, Europhys. Lett. {\bf 52}, 705 (2000)

\bibitem{AndrePRL} A.G. Moreira, R.R. Netz, Phys. Rev. Lett {\bf 87}, 
078301 (2001).

\bibitem{AndreEPJE}
A.G. Moreira, R.R. Netz, Eur. Phys. J. E {\bf 8}, 33 (2002).

\bibitem{Deserno03} M. Deserno, A. Arnold, C. Holm, Macromolecules 
{\bf 36}, 249 (2003). 

\bibitem{Naji_epl04} A. Naji, A. Arnold, C. Holm, R.R. Netz, Europhys. Lett. {\bf 67}, 
130 (2004). 

\bibitem{Lee04} 
K.-C. Lee, I. Borukhov, W.M. Gelbart, A.J. Liu, M.J. Stevens, Phys. Rev. Lett. 
{\bf 93}, 128101 (2004).

\bibitem{Kjellander84} R. Kjellander, S. Mar{\v{c}}elja, 
Chem. Phys. Lett. {\bf 112}, 49 (1984); J. Chem. Phys. {\bf 82},  
2122 (1985).

\bibitem{Oosawa} F. Oosawa, Biopolymers {\bf 6}, 1633 (1968). 

\bibitem{Oosawa_book} F. Oosawa, {\em Polyelectrolytes} (Marcel Dekker, 
New York, 1971). 

\bibitem{Attard-etal} P. Attard, R. Kjellander, D.J. Mitchell, 
Chem. Phys. Lett. {\bf 139}, 219 (1987);  P. Attard, 
R. Kjellander, D.J. Mitchell, B. J\"onsson, J. Chem. Phys. 
{\bf 89}, 1664 (1988). 

\bibitem{Attard-etal2} 
P. Attard, D.J. Mitchell, B.W. Ninham, J. Chem. Phys. {\bf 88}, 4987 (1988).

\bibitem{Podgornik89} R. Podgornik, B. {\v{Z}}ek{\v{s}}, J. Chem. Soc., 
Faraday Trans. II {\bf 84}, 611 (1988); R. Podgornik, J. Phys. A: Math. Gen. {\bf 23}, 
275 (1990). 

\bibitem{Barrat} J.-L. Barrat, J.-F. Joanny, Adv. Chem. Phys. 
{\bf XCIV}, 1 (1996). 

\bibitem{Pincus98} P.A. Pincus, S.A. Safran, Europhys. Lett. 
{\bf 42}, 103 (1998). 

\bibitem{Podgornik98} R. Podgornik, V.A. Parsegian, Phys. Rev. Lett. 
{\bf 80}, 1560 (1998). 

\bibitem{Ha} B.-Y. Ha, A.J. Liu, Phys. Rev. Lett. {\bf 79}, 1289 (1997); 
Phys. Rev. Lett. {\bf 81}, 1011 (1998); Phys. Rev. E {\bf 58}, 6281 (1998); 
Phys. Rev. E {\bf 60}, 803 (1999); Phys. Rev. Lett. {\bf 83}, 2681 (1999);
Europhys. Lett. {\bf 46}, 624 (1999). 

\bibitem{Safran99} D.B. Lukatsky, S.A. Safran, Phys. Rev. E {\bf 60}, 
5848 (1999). 

\bibitem{Kardar99} M. Kardar, R. Golestanian, Rev. Mod. Phys. {\bf 71}, 
1233 (1999).

\bibitem{Netz-orland} R.R. Netz, H. Orland, Eur. Phys. J. E {\bf 1},  
203 (2000).

\bibitem{Ha01} B.-Y. Ha, Phys. Rev. E {\bf 64}, 031507 (2001).  

\bibitem{Lau02} A.W.C. Lau, P. Pincus, Phys. Rev. E {\bf 66}, 041501 
(2002). 

\bibitem{Gonz01}
O. Gonzalez-Amezcua, M. Hernandez-Contreras, P. Pincus, Phys. Rev. E
{\bf 64}, 041603 (2001).

\bibitem{Stevens90} M.J. Stevens, M.O. Robbins, Europhys. Lett. 
{\bf 12},  81  (1990).

\bibitem{Diehl99} A. Diehl, M.N. Tamashiro, M.C. Barbosa, Y. Levin, 
Physica A {\bf 274}, 433 (1999); M.C. Barbosa, M. Deserno, C. Holm, 
Europhys. Lett. {\bf 52}, 80 (2000). 

\bibitem{Rouzina96} I. Rouzina, V.A. Bloomfield, J. Phys. Chem. 
{\bf 100}, 9977 (1996).  

\bibitem{Korny}  A.A. Kornyshev, S. Leikin, J. Chem. Phys. {\bf 107}, 
3656 (1997); Phys. Rev. Lett. {\bf 82}, 4138 (1999). 

\bibitem{Arenzon99} J.J. Arenzon, J.F. Stilck, Y. Levin, Eur. Phys. J. B 
{\bf 12}, 79 (1999); Y. Levin, J.J. Arenzon, J.F. Stilck, Phys. Rev. Lett. 
{\bf 83}, 2680 (1999).

\bibitem{Shklovs99a}
B.I. Shklovkii, Phys.  Rev.  E {\bf 60},  5802  (1999).

\bibitem{Shklovs99} B.I. Shklovskii, Phys. Rev. Lett. {\bf 82}, 3268 
(1999).

\bibitem{Shklovs02}  A. Yu. Grosberg, T. T. Nguyen,  B. I. Shklovskii, 
Rev. Mod. Phys. {\bf 74}, 329 (2002).

\bibitem{Diehl01}  A. Diehl, H.A. Carmona, Y. Levin, Phys. Rev. E
{\bf 64}, 011804 (2001).

\bibitem{Lau01} A.W.C. Lau, D. Levine, P. Pincus, Phys. Rev. Lett. 
{\bf 84}, 4116 (2000); A.W.C. Lau, P. Pincus, D. Levine, H.A. Fertig, 
Phys. Rev. E {\bf 63}, 051604 (2001). 

\bibitem{Netz01} R.R. Netz, Eur. Phys. J. E {\bf 5}, 557 (2001).

\bibitem{Levin02} 
Y. Levin, Rep. Prog. Phys. {\bf 65}, 1577 (2002). 

\bibitem{Naji_epje04}
A. Naji, R.R. Netz, Eur. Phys. J. E {\bf 13}, 43 (2004). 

\bibitem{PhysToday}
W.M. Gelbart, R.F. Bruinsma, P.A. Pincus, V.A. Parsegian, 
Physics Today (September 2000), pp. 38-44.

\bibitem{Rouzina98} I. Rouzina, V.A.  Bloomfield, Biophys. J.
{\bf 74}, 3152 (1998). 

\bibitem{Golestan99} R. Golestanian, M. Kardar, T.B. Liverpool, 
Phys. Rev. Lett. {\bf 82}, 4456 (1999).

\bibitem{Levin99} Y. Levin, Physica A {\bf 265}, 432 (1999).

\bibitem{Shklo00} T.T. Nguyen, I. Rouzina, B.I. Shklovskii, J. Chem. Phys. 
{\bf 112}, 2562 (2000).

\bibitem{Belloni95} O. Spalla, L. Belloni, Phys. Rev. Lett. {\bf 74}, 
2515 (1995); L. Belloni, J. Phys.: Condens. Matter {\bf 12}, R549 (2000). 

\bibitem{Manning97} J. Ray, G.S. Manning, Macromolecules {\bf 30}, 
5739 (1997); Langmuir {\bf 10}, 2450 (1994). 

\bibitem{Neu} J.C. Neu, Phys. Rev. Lett. {\bf 82}, 1072 (1999). 

\bibitem{Sader} J.E. Sader, D.Y.C. Chan, J. Colloid Interface Sci. 
{\bf 213}, 268 (1999); Langmuir {\bf 16}, 324 (2000). 

\bibitem{Trizac} E. Trizac, J.-L. Raimbault, Phys. Rev. E {\bf 60}, 
6530 (1999); E. Trizac, Phys. Rev. E {\bf 62}, R1465 (2000).

\bibitem{Note1}
One can show based on dimensional arguments that the system of counterions at a charged
wall has only two independent length scales ({\em e.g.} the Bjerrum length and the Gouy-Chapman length)
and only one independent dimensionless parameter. 

\bibitem{Baus}
M. Baus, J. Hansen, Phys.\ Rep. {\bf 59},  1  (1980).

\bibitem{Note2}
A more accurate estimate of the typical distance between counterions  at low 
couplings ($\Xi\ll 1$), $a$,  gives $a/\mu \sim \Xi^{1/3}$ taking into account
the 3D structure of the counterionic layer.  

\bibitem{contact_value}
 D. Henderson, L. Blum, J. Chem. Phys. {\bf 75}, 2025 (1981);
S.L. Carnie, D.Y.C. Chan, J. Chem. Phys.  {\bf 74}, 1293 (1981); 
H. Wennerstr{\"o}m, B. J{\"o}nsson,  P. Linse, J. Chem. Phys. {\bf 76}, 
4665 (1982).

\bibitem{Burak04} 
Y. Burak, D. Andelman, H. Orland, Phys. Rev. E {\bf 70}, 016102 (2004).

\bibitem{Andre_thesis}
A.G. Moreira, {\em Charged systems in bulk and at interfaces}, Ph.D. dissertation, Potsdam University
(2001).

\bibitem{Swetlana} 
S. Jungblut, R.R. Netz, unpublished; S. Jungblut, Diploma dissertation, Ludwig Maximilian University
Munich (2004).

\bibitem{Manning69} G.S. Manning, J. Chem. Phys. {\bf 51}, 924 (1969). 

\bibitem{Zimm}
B.H. Zimm, M. Le Bret, J. Biomol. Struct. Dyn. {\bf 1}, 461 (1983);
M. Le Bret, B.H. Zimm, Biopolymers {\bf 23}, 287 (1984).

\bibitem{Naji-Netz-unpub} A. Naji, R.R. Netz, submitted to Phys. Rev. Lett. (2005). 

\bibitem{Note_gammaRB}
Note that $\gamma_{\textrm{RB}}$ may be related to the coupling parameter as 
$\gamma_{\textrm{RB}}=\Xi/(\xi\tilde \Delta)$, where $\tilde \Delta=\Delta/\mu$. 

\bibitem{DBL}
A. Diehl, M.C. Barbosa, Y. Levin, Europhys. Lett. {\bf 53}, 86 (2001). 

\bibitem{Andre02} 
A.G. Moreira, R.R. Netz, Europhys. Lett. {\bf 57}, 911 (2002).

\end{thebibliography}
\end{document}